\documentclass[aps,reprint,groupedaddress,nobibnotes,longbibliography]{revtex4-1}

\usepackage{graphicx}
\usepackage{natbib}
\usepackage{amsmath,amssymb}
\usepackage{subfig}
\usepackage{setspace}
\usepackage{float}
\usepackage{microtype}
\usepackage[normalem]{ulem}

\usepackage[usenames,dvipsnames]{color}
\usepackage{mathtools}
\usepackage{ragged2e}
\DeclareCaptionJustification{justified}{\justifying}
\usepackage{hyperref}
\hypersetup{
	hyperindex,
	breaklinks,
	colorlinks=true,
	linkcolor=blue,
	citecolor=magenta,
	bookmarks=true,
	bookmarksopen=true,
	bookmarksopenlevel=2,
	pdfstartpage={1},
	pdfstartview={FitH},
	pdfview={FitH 0},
	pdfauthor={B. F. Farrell and P. J. Ioannou},
	pdftitle={Statistical state dynamics-based analysis of the physical mechanisms sustaining and regulating turbulence in Couette flow}}

\usepackage{ifthen}

\newcommand{\df}{\textrm{d}}

\newcommand\Idm{\mathbf{I}}

\newcommand\Mm{\mathbf{M}}

\newcommand\Lm{{\mathbf{L}}}

\newcommand\Cm{{\mathbf{C}}}
\newcommand\Sm{{\mathbf{S}}}
\newcommand\Dm{{\mathbf{D}}}

\newcommand\Am{{\mathbf{A}}}
\newcommand\Em{{\mathbf{E}}}
\newcommand\Um{{\mathbf{U}}}
\newcommand\Qm{{\mathbf{Q}}}
\newcommand\Fm{\mathbf{F}}

\renewcommand\Lm{{\mathbf{L}}}
\renewcommand{\(}{\left(}
\renewcommand{\)}{\right)}
\renewcommand{\[}{\left[}
\renewcommand{\]}{\right]}

\def\bit{\vphantom{\dot{W}}}

\def\bU{\boldsymbol{U}}
\def\bu{\boldsymbol{u}}

\def\uv{\bu}

\providecommand\bcdot{\boldsymbol{\cdot}}
\providecommand\bnabla{\boldsymbol{\nabla}}
\providecommand\Phim{\boldsymbol{\Phi}}

\newboolean{showcomments}
\setboolean{showcomments}{true}

\newcommand{\dennice}[1]{  \ifthenelse{\boolean{showcomments}}
{\textcolor{blue}{(Dennice says:  #1)}}{}}
\newcommand{\brian}[1]{\ifthenelse{\boolean{showcomments}}
{\textcolor{Red}{(Brian says: #1)}}{}}
\newcommand{\petros}[1]{\ifthenelse{\boolean{showcomments}}
{\textcolor{Green}{Petros says: #1)}}{}}
\newcommand{\vaughan}[1]{\ifthenelse{\boolean{showcomments}}
{\textcolor{Purple}{Vaughan says: #1)}}{}}

\def\i{\textrm{i}}

\begin{document}

\title{Statistical state dynamics-based analysis of the physical mechanisms sustaining and regulating turbulence in Couette flow}

\author{Brian F. Farrell}
\affiliation{School of Engineering and Applied Science, Harvard University}

\author{Petros J. Ioannou}
\email{pjioannou@phys.uoa.gr}
\affiliation{Department of Physics, National and Kapodistrian University of Athens}

%

\date{\today}

 \begin{abstract}
 This paper describes a study of the self-sustaining process  in  wall-turbulence.
 The study  is based on  a second order statistical state dynamics  model of Couette flow
 in which the state variables are  the streamwise mean  flow (first cumulant) and perturbation covariance  (second cumulant).  This statistical state dynamics model
is closed  by either setting the third cumulant to zero or by replacing it with a stochastic parameterization.
Statistical state dynamics models with this form are referred to as  S3T models.  S3T models have been shown to
self-sustain turbulence with a mean flow and second order perturbation structure  similar to that obtained by direct numerical
simulation of the equations of motion. The use of a statistical state dynamics model to study the physical mechanisms
underlying turbulence
has  important advantages over the traditional approach of studying
the dynamics of individual realizations of turbulence.   One advantage is that the analytical structure of S3T statistical state dynamics models isolates the interaction between the mean flow and the
perturbation components of the turbulence.  Isolation of the  interaction between these components reveals how this interaction underlies both the maintenance of the turbulence variance by transfer of energy from the externally driven flow to the perturbation components as well as the enforcement of the observed statistical mean turbulent state by feedback regulation  between the mean and perturbation fields. Another advantage of studying turbulence using statistical state dynamics models of S3T form
is  that the analytical structure of S3T turbulence can be completely characterized. For example, the perturbation component of turbulence in the S3T system is demonstrably maintained by a parametric perturbation growth mechanism in which  fluctuation of
the mean flow maintains the perturbation field which in turn maintains
the mean flow fluctuations in a synergistic  interaction. Furthermore, the equilibrium statistical state of S3T
turbulence can be demonstrated to be enforced by feedback regulation in which transient growth of the
perturbations episodically suppresses
streak growth preventing runaway parametric growth of the perturbation component. Using S3T to isolate  the
parametric growth and feedback regulation mechanisms allows a detailed characterization of the dynamics of the  self-sustaining process in S3T turbulence with compelling implications for advancing understanding of wall-turbulence.

\end{abstract}


\maketitle

\section{Introduction}

In this work  the mechanisms  sustaining and regulating wall-turbulence are studied.
Understanding how the turbulent state is sustained against dissipation requires identifying the mechanism by which energy is systematically
transferred from the externally driven  flow to the turbulent fluctuations in the absence of fast inflectional instability of the mean velocity profile.
Understanding how the  turbulence is enforced to assume the observed statistical structure
requires also understanding the mechanism by which interaction between the mean and perturbation fields establishes and enforces this observed statistical state.

The ubiquitous
roll-streak structure, which was first identified in the buffer layer~\cite{Kline-etal-1967}, is known to play a key role
in the dynamics of wall-turbulence.
While the roll-streak structure is stable in plane wall-bounded flows, it
produces robust energy transfer
from the mean shear flow to the perturbation field when an optimally configured perturbation is excited~\citep{Butler-Farrell-1992, Reddy-Henningson-1993}.
This growth results from a streamwise roll circulation giving rise to  a streamwise  streak  through the lift-up mechanism~\cite{Landahl-1980}.  What is not understood is how this mechanism is maintained in the absence of a linear instability.
An early proposed resolution of this conundrum was that these structures participate in a regeneration cycle in which  new streaks arise from perturbations the origin of which is ascribed to the break-up of previously formed  streaks~\cite{Swearingen-Blackwelder-1987, Blakewell-Lumley-1967}.
This proposed cycle  can be viewed as a nonlinear instability mechanism in which turbulence is sustained by energy transfer from the externally forced shear to the perturbation field due to the linear non-normal lift-up growth processes while this non-normal growth is in turn sustained nonlinearly through the continual re-emergence from streak breakdown
 debris of perturbations configured to excite subsequent roll-streak structures.
Alternative mechanisms by which perturbation nonlinearity could sustain transient growth of the roll-streak structure
have been the subject of  study since this nonlinear basis of the instability maintaining wall-turbulence was postulated.  Here we refer to these mechanisms collectively as self-sustaining processes.

An alternative to the regeneration cycle class of self-sustaining process is based on spanwise inflection of the streak velocity profile giving rise to an unstable, or at statistical equilibrium a neutral, eigenmode.
In this class of self-sustaining process    Reynolds stresses arising from this eigenmode  sustain the roll circulation~\cite{Waleffe-1995a,Hamilton-etal-1995,Waleffe-1997,Hall-Sherwin-2010}.  However, subsequent work indicated that streaks are often too weak to be  unstable and  an alternative self-sustaining process was postulated in which transient
growth rather than modal instability maintains the perturbations that force the roll~\cite{Schoppa-Hussain-2002}.    An advantage of the transient growth self-sustaining process is that the optimal perturbations  maximally
exploit the energy of the wall-normal shear in addition to the spanwise shear that primarily supports the  inflectional instability.  In fact, the most rapidly growing perturbations in shear flow are oblique waves which optimally exploit, by lift-up, the large reservoir of energy in the wall-normal shear~\cite{Farrell-Ioannou-1993a}.   Moreover, the Reynolds stresses arising in association with these optimally growing
oblique waves  have been shown to give rise
to the strong systematic forcing of the roll circulations required to maintain
the streak~\cite{Farrell-Ioannou-2012} and consistently,
oblique waves are commonly observed to accompany streaks in wall-turbulence ~\cite{Schoppa-Hussain-2002}.

Insight into the self-sustaining process was advanced by recent work in which it was shown that roll-streak formation  is not confined to the boundary layer as had been  previously established \cite{Jimenez-Moin-1991, Hamilton-etal-1995,
Jimenez-Pinelli-1999} but is operating similarly throughout the shear flow \cite{Farrell-etal-2016-VLSM, Hwang-Cossu-2010b,Hwang-Cossu-2011,Cossu-Hwang-2017}.
This suggests a universal underlying self-sustaining process  mechanism in shear flow that is not scale selective.
Such a mechanism was identified to be the streak amplification process  resulting from the organization of supporting roll circulation by perturbation scale turbulence \cite{Farrell-Ioannou-2012}. Because this universal mechanism is not scale selective,
scale restriction, such as that  imposed by Hwang and Cossu \cite{Hwang-Cossu-2010b,Hwang-Cossu-2011,Cossu-Hwang-2017},
need only include scales nearby the scales of the selected roll-streak in order to include the  oblique waves and associated adjoint perturbations  which support the self-sustaining process at a selected roll-streak scale.

While these various self-sustaining process mechanisms address the question of how the roll-streak
structure might be nonlinearly destabilized, they leave open the question of how this instability is regulated to
zero mean growth and more generally how the turbulence is enforced to assume
the observed statistical equilibrium state. Both of these questions can be addressed
using a statistical state dynamics
model~\citep{Farrell-Ioannou-2012,Farrell-etal-2016-PTRSA}. While analyzing complex spatially and temporally varying fields arising in observations and simulations of turbulent systems using statistical quantities is common practice, it is less
 common to adopt statistical  variables  directly as the variables for expressing the dynamics of the turbulent system.  An early attempt to exploit the potential of
 employing statistical state dynamics  to provide insight into turbulence involved formal
 expansion and closure of the Navier-Stokes equations in cumulants~\cite{Hopf-1952,Kraichnan-1964, Frisch-1995}.
 The cumulant method was subsequently restricted in application in  part due to the difficulty
of obtaining robust closure of the expansion when it was applied to isotropic homogeneous turbulence.
Surprisingly,  while  the assumed
vanishing of the first cumulant in isotropic homogeneous turbulence would appear to  simplify the
dynamics, subsequent
experience in solving statistical state dynamics models in the cases of  anisotropic  two dimensional
beta-plane turbulence~\cite{Farrell-Ioannou-2007-structure,Srinivasan-Young-2012,Tobias-Marston-2013,Constantinou-etal-2014} and turbulent convection~\cite{Herring-1963,Herring-1964}
revealed that closures retaining nontrivial expressions for the first and in addition only the second cumulant comprise the entire essential dynamics of the turbulence. For example, statistical state dynamics of beta-plane turbulence closed at second order while retaining the streamwise mean as the first order cumulant predicts the equilibrium state of this turbulence to be an analytical solution (in the form of a fixed point) of the statistical state dynamics including the remarkable spontaneous formation of jets with the observed
structure containing as much as $90\%$ of the kinetic energy of the flow~\cite{Farrell-Ioannou-2003-structural,Farrell-Ioannou-2007-structure, Farrell-Ioannou-2009-closure,Farrell-Ioannou-2009-equatorial,Srinivasan-Young-2012, Parker-Krommes-2013-generation, Bakas-Ioannou-2013-prl, Tobias-Marston-2013}. In retrospect, precedence for such a program was provided by the work of Herring~\cite{Herring-1963,Herring-1964} in his study of the  statistical equilibrium of turbulent convection.
The approach of using second order statistical state dynamics to obtain the statistical equilibrium state of turbulent
convection has its roots in Malkus's theory in which the statistical state was sought as the fixed point equilibrium between the mean thermal structure  and the turbulent heat fluxes~\cite{Malkus-1954, Malkus-Veronis-1958}.
The success of this program in providing an explanation for the statistical mean state of turbulent convection was aided by the underlying instability being a temporal normal mode which could be equilibrated by second order thermal fluxes (obtained from the second cumulant) modifying  the time-independent thermal structure of the mean state (obtained from the first cumulant) to stability.
The successful application of statistical state dynamics  to turbulent convection
encouraged a program of applying the statistical state dynamics approach to understand the dynamics of  anisotropic 3D wall-bounded turbulence.
However, attempts to extend the program of Malkus to obtain the equilibrium state of wall-turbulence as the fixed point of a second order
closure of the statistical state dynamics  did not succeed~\cite{Reynolds-Tiederman-1967}.
From the point of view presented in this work the concept of applying the  program of Malkus~\cite{Malkus-1956} to wall turbulence was essentially correct requiring only the additional recognition that the instability  to be equilibrated is the instability of the
time-dependent operator associated with linearization about the temporally varying streamwise mean flow.
In contrast, the program of Malkus and its variations was predicated on
stabilizing the temporal modal instability associated with linearization about the time-independent mean flow.
With the additional insight that the instability maintaining the perturbation variance  in shear flow turbulence is
parametric the equilibrium turbulent state  is understood to result from quasi-linear
 adjustment of the time-dependent mean flow  to neutral parametric growth rate of its most unstable structure  or structures.
The growth  rate of these temporally varying structures is given by the  maximal Lyapunov exponent
of the perturbation covariance equation (this growth rate is necessarily zero given that the turbulence statistics are stationary).

As remarked by  Herring~\cite{Herring-1963} second order closures of the statistical state dynamics are necessarily quasi-linear.
SSD models that use  a finite ensemble approximation to  estimate the second cumulant
in the statistical state  dynamics
of the Navier--Stokes equations
are  consistently also   quasi-linear and  we  refer to such systems as RNL$_N$ systems (restricted nonlinear systems of order $N$).
The simplest RNL$_N$ system is the RNL$_1$ system which consists of the streamwise mean  equations
forced by the Reynolds stresses  obtained from the perturbation covariance formed using
a single realization of the perturbation dynamics.  While the dynamics of the RNL$_1$ system is formally equivalent to that of a quasi-linear system  consisting of a mean flow and a realization of the perturbation dynamics, RNL$_N$
systems for $N>1$ can only be regarded as approximation to the second order statistical state dynamics.  Consistently, we regard our state variables to be the mean flow and the covariance of the perturbations (the first and second cumulants) regardless of how many ensemble members are used to approximate the covariance.

As $N$ increases RNL$_N$
systems  approach  S3T dynamics,  which is a closure of
the statistical state dynamics at second order in which an equivalently infinite ensemble is solved for  by
using a time dependent
Lyapunov equation~\citep{Farrell-Ioannou-2003-structural}.  The S3T system is closed
by either setting the third cumulant to zero or by replacing it with a stochastic parameterization.
Because S3T dynamics is recovered in the limit $N\to\infty$
we can identify solutions of RNL$_\infty$ systems with S3T. While the use in
S3T dynamics of a time-dependent Lyapunov equation to advance the perturbation covariance in time allows direct solution for the second cumulant
corresponding to the covariance obtained from a formally infinite ensemble, the great advantage of  RNL$_N$ systems is in allowing extension of second order S3T statistical state dynamics methods to study turbulence at high Reynolds number~\citep{Constantinou-etal-Madrid-2014,Bretheim-etal-2015,Farrell-etal-2016-VLSM}.

The self-sustaining process operating in the S3T  system is similar in some ways to previously proposed self-sustaining processes in that a quasi-linear interaction occurs between perturbations and the mean flow to maintain the roll  which in turn forces the streak completing the cycle of nonlinear instability~\cite{Farrell-Ioannou-2012}.
For example, the self-sustaining process of Waleffe~\cite{Waleffe-1997}
and vortex-wave interaction process of Hall \& Sherwin and collaborators~\citep{Hall-Smith-1991,Hall-Sherwin-2010,Deguchi-etal-2013} are  also
quasi-linear,
although this quasi-linearity is imposed by construction  rather than resulting
from a closure.  
In the self-sustaining process of Waleffe and that of  vortex-wave interaction   a single unstable or neutral inflectional mode
interacts  with the mean flow to transfer energy from the streak to maintain the roll.
In contrast, in S3T the streamwise mean flow interacts with a broad spectrum of
background turbulence in producing the energy transfer that maintains the roll by
a fundamentally non-modal process. Moreover, unlike previously proposed modal instability-based mechanisms or transient growth-based mechanisms~\cite{Jimenez-2013,Schoppa-Hussain-2002}, the growing perturbations sustaining S3T turbulence result from parametric instability of the time-dependent streak~\cite{Farrell-Ioannou-2012}. Parametric instability
is generally associated with its application to the study of the stability of a periodically modulated system (cf.~Drazin~\&~Reid~\cite{Drazin-Reid-81}, section 48).
We widen application of this term to refer to any linear instability that
is inherently caused by the time dependence of the system. The reason we have adopted the same word to describe the instability of periodic and non-periodic
flows is that the same  non-normality based instability mechanism operates
in both cases~\cite{Farrell-Ioannou-1996b, Farrell-Ioannou-1999}.
 Moreover, analysis reveals that this  parametric instability mechanism
destabilizes all  linear time-dependent dynamical systems that
fluctuate with sufficiently high amplitude~\cite{Farrell-Ioannou-1999}.
This  instability almost surely manifests asymptotically in time
in dominance of the
perturbation dynamics by the structure of the  top Lyapunov vector  (or vectors in the case of degeneracy)
of the associated time dependent linear dynamical operator~\cite{Oseledets-1968, Lorenz-1984, Farrell-Ioannou-1996b}.
%

In previous work we identified the parametric instability mechanism
underlying maintenance of the perturbation variance in S3T Couette turbulence~\cite{Farrell-Ioannou-2012,Thomas-etal-2015}.    We also noted
its  association with the  theory of the instability of stochastic
time-dependent linear dynamical systems. It follows from this theory that the perturbation dynamics can be decomposed into a
basis of  Lyapunov vectors
each characterized
by a Lyapunov exponent~\cite{Oseledets-1968,Lorenz-1984}.
Time dependent systems are non-normal with measure zero exception~\citep{Farrell-Ioannou-1996b,Farrell-Ioannou-1999}
and there is an analogy between the Lyapunov vectors of a time
dependent dynamical system and the familiar example  of the eigenvectors of a  time independent non-normal linear
system. In both of these cases the dynamics can be expressed using  a basis of eigenvectors each  characterized
by the exponential growth rate of its associated eigenvalue.
This analogy suggests a program of exploiting the known analytical Lyapunov structure
of the parametric stability of random time-dependent dynamical
systems to gain insight into the dynamics of the perturbation component of
the turbulence and particularly the mechanism  that maintains  it.
Heretofore this perturbation component   has been generally thought of as
resulting from random transient growth events and scattering by
perturbation-perturbation nonlinearity. Characterization of the perturbation
component of the turbulence in terms of Lyapunov vectors offers the possibility of understanding
the maintenance and structure of the perturbation component of turbulence in
wall-bounded shear flows
 more precisely.  A realistic turbulence exists naturally in the S3T
 self-sustaining state for Couette flow in which  only the first Lyapunov vector is
 supported~\citep{Farrell-Ioannou-2012,Thomas-etal-2014,Bretheim-etal-2015,Farrell-etal-2016-VLSM}
 providing  complete characterization of this turbulence.
 The closest  analogue to the eigenvectors of a time independent non-normal
 operator  are the Lyapunov vectors of Oseledets~\citep{Oseledets-1968}, which are referred to as the confluent
 Lyapunov vectors (CLV) \cite{Ginelli-etal-2007,Wolfe-Samelson-2007,Yang-Radons-2012,Cvitanovic-etal-2016}.
 However, as we are interested primarily in perturbation energetics it suffices for our purposes to work with the more easily calculated set of related Lyapunov vectors introduced by Lorenz \citep{Lorenz-1984}. These Lyapunov vectors (LV's) correspond to  orthogonalization of the CLV's in the energy norm.
 The  LV's span the same perturbation state space and have the same Lyapunov exponents as the CLV's but they have been
 rotated in the spanned space so as to be orthogonal in the energy norm.
 The relation between the CLV's and the LV's is further discussed in Appendix ~\ref{ap:A}.

In this work we isolate the instability mechanism supporting the
perturbation structure from the turbulence dynamics  by obtaining the time-dependent mean flow from a self-sustaining S3T turbulence and using  this time dependent mean flow to force the instability of a completely separate perturbation dynamics that has been randomly initialized.
This program is analogous to taking an inflectional streak from an observation of a stationary shear flow and calculating
the most unstable temporal normal mode on this streak: one would predict the form of the perturbation structure to be that of the
fastest growing mode.
 In complete analogy we can  predict the structure of the turbulent perturbations in this S3T Couette turbulence to be that of the first
Lyapunov vector perturbation on  the corresponding time-dependent mean flow.
Having obtained the structure of the perturbation component of this simplified turbulence  we then proceed to
characterize it in terms of its energetics and mechanism of growth.
Having obtained complete characterization of this simplified turbulence supported by only the first
Lyapunov vector we then proceed to study the energetics of the remaining Lyapunov vectors which, although damped in this simple model with no parameterized nonlinearity, are expected to be maintained at finite amplitude by scattering arising from perturbation nonlinearity when a parameterization for nonlinearity is included.
The result  of implementing such a parameterization for perturbation--perturbation nonlinearity  is the
prediction that the remaining Lyapunov vectors  are
robustly supported by direct energetic interaction with the time dependent mean flow.   The implication of this result is
contrary to the idea that the perturbation variance
in turbulent shear flow
results from a  cascade or from random transient growth events
suggesting  rather that
spectrally nonlocal interaction with the fluctuating mean flow  constitutes a primary mechanism for
maintaining the perturbation variance.  Consistently, we show that  these analytically
characterized Lyapunov vectors together comprise the dominant support for
 the perturbation variance structure.
Having understood the perturbation dynamics in
isolation we next proceed to recouple the mean and
perturbation systems to recover the complete S3T turbulence dynamics and use this system to study the feedback control mechanism that
regulates the turbulence to its statistical steady state.

\section{The S3T statistical state dynamics model}
\label{sec:framework}

Consider  plane Couette flow  between walls with velocities $\pm U_w $.
The streamwise direction is $x$, the wall-normal direction is $y$, and the spanwise direction is $z$.
Lengths are non-dimensionalized by the channel half-width, $\delta$, and
velocities by $U_w$, so that  the Reynolds number is $Re= U_w \delta / \nu$, with $\nu$ the coefficient of kinematic viscosity.
We take for our example a doubly periodic channel of non-dimensional length $L_x$ in the streamwise direction and $L_z$ in the spanwise.

The  velocity field is  decomposed into a streamwise mean, $\bU$, with components, $(U,V,W)$, and
perturbation from this mean, $\uv$,  with components $(u,v,w)$.
The pressure is similarly decomposed  into its streamwise mean, $P$, and  perturbation from this mean, $p$.
The non-dimensional Navier-Stokes equations decomposed into an equation for the streamwise mean and an equation for the perturbation are:
\begin{subequations}
\label{eq:NS}
\begin{gather}
\partial_t\bU+ \bU \bcdot \bnabla \bU   + \bnabla P -  \Delta \bU/R = - \[\bu\bcdot \bnabla \uv\]_x\ ,
\label{eq:NSm}\\
\partial_t\uv+   \bU \bcdot \bnabla \bu +
\bu \bcdot \bnabla \bU  + \bnabla p-  \Delta  \uv/R
=  \boldsymbol{\mathcal{N}}~,
 \label{eq:NSp}\\
 \bnabla \bcdot \bU = 0\ ,\ \ \ \bnabla \bcdot \uv = 0\ \ , \label{eq:NSdiv0}
\end{gather}\label{eq:NSE0}\end{subequations}
where $\boldsymbol{\mathcal{N}} \equiv  \[\bu \bcdot \bnabla \bu\]_x-\bu \bcdot \bnabla \uv  $ is the contribution
to perturbation dynamics from
the perturbation-perturbation interactions. Square brackets denote an average over the variable that appears as subscript, e.g.
\begin{equation}
[\,\bcdot\,]_x \equiv L_x^{-1} \int_0^{L_x} \bcdot\ \df x\ ,
\label{eq:x}
\end{equation}
is a streamwise average and $[\,\bcdot\,]_{x,z}$ a streamwise and spanwise average.
The velocities  satisfy periodic boundary
conditions in the $z$ and $x$ directions and no-slip boundary
conditions in the wall-normal  direction: $\bU(x,\pm 1,z,t)= (\pm 1,0,0)$, $\bu (x,\pm1,z,t)=0$.

The statistical state dynamics of~\eqref{eq:NS} is closed at second order by
parameterizing the
perturbation--perturbation interactions,  $\boldsymbol{\mathcal{N}}$,
as a  nondivergent stochastic excitation and associated dissipation, $\boldsymbol{\mathcal{G}}$,
chosen to satisfy at every time instant:
\begin{equation}
 \[ \bu \cdot   \boldsymbol{\mathcal{G}}  \]_{x,y,z} = 0~,\label{eq:cond}
\end{equation}
consistent with the requirement that
perturbation--perturbation interactions redistribute
energy among the  perturbation components without introducing  any net energy into the perturbation field.

 With this parameterization the Navier-Stokes equations are reduced to  this
 quasi-linear equation set:
 \begin{subequations}\label{eq:RNL}\begin{gather}
\partial_t\bU+ \bU \bcdot \bnabla \bU   + \bnabla P -  \Delta \bU/R= - \[\bu\bcdot \bnabla \uv\]_x\ ,
\label{eq:RNLm}\\
 \partial_t\uv+   \bU \bcdot \bnabla \bu +
\bu \bcdot \bnabla \bU  + \bnabla p-  \Delta  \uv/R
= \boldsymbol{\mathcal{G}} ~.\label{eq:RNLp}
\end{gather}\end{subequations}

If the perturbation covariance obtained from this
quasi-linear RNL$_1$
system is regarded as an approximation to the second cumulant an approximate closure of the related
second order SSD is obtained.  This closure supports realistic
turbulence and has proven useful in studying turbulence dynamics~\cite{Thomas-etal-2014,Thomas-etal-2015,
Bretheim-etal-2015}. An $N$-member ensemble of
independent perturbation systems of
form~\eqref{eq:RNLp} sharing the same mean flow $\bU$ solving~\eqref{eq:RNLm}
provides an approximation to the statistical state dynamics of the S3T system
referred to as RNL$_N$~\cite{Farrell-etal-2016-PTRSA}.

It is convenient  to use the non-divergence of the mean flow  to
express the mean dynamics~\eqref{eq:RNLm} in terms of the mean streamwise velocity,  $U$, and a mean spanwise/wall-normal
velocity streamfunction, $\Psi$. In these variables~\eqref{eq:RNLm} is equivalent to
\begin{subequations}\label{eq:MNSE}\begin{align}
&\partial_t U = U_y \Psi_z - U_z \Psi_y -\partial_y  \[ uv \]_x  -\partial_z \left [ uw\right ]_x + \Delta_1 U/R,\label{eq:MU}\\
&\partial_t \Delta_1 \Psi =  (\partial^2_{y}-\partial^2_{z})\left (\Psi_y \Psi_z  - \left [ vw \right ]_x \right ) \nonumber \\
& \hspace{.2em}- \partial_{yz}\left ( \Psi_y^2 - \Psi_z^2 + \left [ w^2 \right ]_x - \left [ v^2 \right ]_x  \right ) +  \Delta_1 \Delta_1 \Psi /R,
\label{eq:MPSI}\end{align}\end{subequations}
with $\Delta_1\equiv \partial_{y}^2+\partial_{z}^2$ and the mean wall-normaland spanwise velocities are given by $V=-\Psi_z$ and $W=\Psi_y$ respectively. Subscripts in flow fields denote differentiation in the variable indicated by the subscript.

Nondivergence of the perturbation velocity field is used to
eliminate the pressure from the perturbation equations~(\ref{eq:RNLp}) by transforming the perturbation dynamics into the variables
wall-normal velocity $v$ and wall-normal vorticity,
$\eta\equiv u_z - w_x$. In these variables equations~\eqref{eq:RNLp},
upon neglect of the advection of the perturbations by the smaller magnitude $V$ and $W$ velocities (i.e. by neglecting
terms $V \partial_y \bu$, $W \partial_z \bu$, $\bu \bcdot \bnabla V$, and $\bu \bcdot \bnabla W$ in~\eqref{eq:RNLp}),
assume the convenient form:\begin{subequations}\label{eq:PNSE}\begin{align}
&\partial_t \Delta v +   U  \Delta v_x + U_{zz} v_x  + 2 U_z v_{xz} - U_{yy} v_x - 2 U_z w_{xy}
\nonumber \\
& \hspace{2em}-  2 U_{yz} w_x - \Delta \Delta v/R  = {\mathcal{G}}_v~,  \label{eq:pv} \\
&\partial_t \eta  +  U \eta_x  - U_z v_y + U_{yz} v + U_y v_z+ U_{zz} w   - \Delta \eta/R =\nonumber \\
&  \hspace{2em} = {\mathcal{G}}_\eta ,\label{eq:peta}
\end{align}\end{subequations}
where $ {\mathcal{G}}_v$ and $ {\mathcal{G}}_{\eta}$ is the parameterization of the perturbation-perturbation interactions  in these variables.

We next Fourier expand the perturbation fields and the stochastic excitation fields in $x$, e.g:
 \begin{equation*}
 	v= {\rm Re} \left[\vphantom{\dot{W}}\right. \sum_{k>0} \hat v_{k}(y, z, t) \,\textrm{e}^{\i  k x}\left.\vphantom{\dot{W}}\right]\ ,
\end{equation*}
where $\rm Re$ denotes the real part, and then write the equations~(\ref{eq:PNSE})  for the evolution of the Fourier components
 of the perturbations in  matrix form:
\begin{equation}
\frac{ d \tilde{\phi}_k }{ d t}
= \tilde{\Am}_k({ U}) \tilde{\phi}_k +   \, \tilde{\Fm}_{k} \xi(t) - r \tilde{\phi}_k\ ,\label{eq:Aphit}
\end{equation}
where the state of the system $\tilde{\phi}_k = [\hat{v}_k,\hat{\eta}_k]^T$  comprises the values of the $\hat{v}_k$ and $\hat{\eta}_k$ on the $N=N_y N_z$
grid points of the $(y,z)$ plane.  The Fourier amplitudes of the perturbation fields satisfy periodic boundary conditions in $x$ and $z$ and $\skew2\hat{{v}}_k= \partial_y \skew2\hat{{v}}_k =\skew2\hat{{\eta}}_k=0$ at $y=\pm1$. The matrix
$\tilde{\Am}_k$ is the discretized Orr--Sommerfeld and Squire operator
for perturbations with $x$-wavenumber $k$  evolving about  the instantaneous
mean streamwise flow
$U(y,z,t)$~\cite{Schmid-Henningson-2001,Farrell-Ioannou-2012}.
We have parameterized
${\boldsymbol{\cal G}}$, as   $\tilde{\Fm}_{k} \xi(t) - r \tilde{\phi}_k$
where $\tilde{\Fm}_k$  is the $2 N \times 2 N$ matrix determining the spatial structure of the stochastic excitation, $r$ is a linear dissipation coefficient and $\xi$ a $2 N$ vector of independent zero-mean stochastic processes satisfying:
\begin{equation}
	\langle \xi (t_1)\xi(t_2)^\dagger \rangle= \delta(t_1-t_2)~\Idm~,
\end{equation}
where $\Idm$ is the $2N$-identity matrix and $\dagger$ denotes the Hermitian transpose. The spatial structures of the forcing,~$\tilde{\Fm}_k$,  do not affect the dynamics so long as the set $\tilde{\Fm}_k$
forms a complete basis for the  forcing  in the $y-z$ plane \citep{Farrell-Ioannou-1993e}. The forcing is chosen to be white in energy and is expressed using  the complete basis consisting of Fourier modes in $z$  and the eigenmodes of the Orr--Sommerfeld and Squire operator in $y$. The specific choice of the basis  was made in order to satisfy the boundary conditions.

The energy density of the perturbations is given by $E= \tilde{\phi}_k^\dagger \Mm_k \tilde{\phi}_k$, where $\Mm_k$ is the
energy metric. It is convenient to consider the perturbation dynamics~\eqref{eq:Aphit} transformed to generalized velocity coordinates
$\phi_k =\Mm_k^{1/2} \tilde{\phi}_k$, so that energy is given by the~$L_2$ norm~$E= \phi_k^\dagger \phi_k$.
The perturbation dynamics in generalized velocity coordinates are governed by:
\begin{equation}
\frac{ d \phi_k }{ d t}
= {\Am}_k({ U}) \phi_k +   \, \Fm_{k} \xi(t)-r {\phi}_k ~,\
\label{eq:Aphi}
\end{equation}
where $\Am_k = \Mm_k^{1/2} \tilde{\Am}_k \Mm_k^{-1/2}$ and $\Fm_k = \Mm_k^{1/2} \tilde{\Fm}_k$.
The linear dissipation  rate $r$ is chosen so that no net energy is introduced by the stochastic excitation
consistent with condition \eqref{eq:cond} being satisfied at each time instant (and   at every $k$).
Delta correlation in time
also implies that the mean energy input by the excitation
is independent of the flow state.

The parameterization for perturbation-pertubation nonlinearity, ${\boldsymbol{\cal G}}$, is highly simplified in order to probe the perturbation dynamics in the least ambiguous manner.  First, this parameterization introduces no energy so any perturbation variance is clearly not being supported by the excitation itself as would be the case e.g. for a stochastically forced pendulum. Second, this parameterization excites each degree of freedom equally so no structural bias is introduced into the energetics as would be the case e.g. a scale dependent excitation were used.  If the dynamics were normal and stable no perturbation variance would be maintained by this parameterization.  Any variance maintained by this parameterization in the case of a non-normal operator arises from induced transfer of energy from the mean flow to the perturbations rather than from the excitation itself.  This parameterization
isolates the proposed mechanism for maintaining perturbation
variance in wall turbulence: parametric transfer directly from the mean
flow to the perturbation field and in particular primarily to the Lyapunov structures.

For an equation of form~\eqref{eq:Aphi} the ensemble average perturbation covariance, $\Cm_k = \langle \phi_k \phi_k^{\dagger} \rangle$,
can be verified to evolve according  to the time-dependent Lyapunov equation:
\begin{equation}
  \frac{{d} \Cm_k }{{d} t}~ = \Am_k( {U} ) \,\Cm_k +\Cm_k \,\Am_k^{\dagger}({ U} ) +  \Qm_k - r \Cm_k~,
   \label{eq:Lyap1}
  \end{equation}
in which: $\Qm_k= \Fm_k  \Fm_k^{\dagger}$~\cite{Farrell-Ioannou-1996a,Farrell-Ioannou-2003-structural}.
The required linear damping is
\begin{equation}
r =\frac{\sum_k{\rm Tr}(\Qm_k)}{\sum_k{\rm Tr}(\Cm_k)}~,
\end{equation}
with $\rm Tr(\,\bcdot\,)$ denoting the trace,  $\sum_k {\rm Tr}( \Cm_k)$ the total perturbation energy
and $\sum_k {\rm Tr}{(\Qm_k)}$ the net energy input rate to all  wavenumbers by the stochastic excitation.
With this  choice for the linear damping  the net energy input  rate
is equal to the perturbation energy dissipation rate at each time instant and no net energy is input to the perturbation field.
A similar parameterization was previously used to close a statistical state dynamics model of baroclinic turbulence~\cite{Farrell-Ioannou-2009-closure}.

The linear equation \eqref{eq:Lyap1}
can be interpreted as the transport equation for the turbulent Reynolds stresses \cite{Hanjalic-1972}
with the first and second  term on the RHS comprising the linear terms expressing convection, generation, destruction, redistribution and diffusion,  while the third and fourth term parameterizes the nonlinear  component of the diffusion and destruction.

Finally, we note that under  the ergodic assumption that
streamwise averages are equal to ensemble averages  the Reynolds stress divergences appearing in the streamwise mean equations~\eqref{eq:MNSE} can be expressed as a linear function of the ensemble average $\Cm_k$ obtained from the time-dependent Lyapunov equation.

With the parameter choice of our example problem  S3T turbulence self-sustains
by interaction between the single perturbation structure with
wavenumber $k= 2 \pi /L_x$ and the mean flow.  Perturbations supported by other streamwise wavenumbers that
happen to be present in an initial state can be verified to have negative Lyapunov
exponents and therefore  damp out in the absence of explicit
excitation at these other wavenumbers and are not retained in the solution  \cite{Farrell-Ioannou-2012, Thomas-etal-2015,
 Farrell-etal-2016-VLSM}. Further,  because in the S3T equations
the streamwise wavenumber perturbation--perturbation interactions are not
retained  there is no mechanism by which energy can enter or
be maintained in streamwise wavenumbers other than the wavenumbers  that are either externally excited or naturally
maintained by the parametric mechanism, which in our case is only $k=2\pi/L_x$.
Consequently, because a single $k$ is retained in the perturbation dynamics
the
subscript on $k$  in the velocity and excitation components is dropped without ambiguity.
The S3T system so restricted self-sustains turbulence in minimal channel Couette flow even at $R=400$~\cite{Farrell-Ioannou-2012}.

Summarizing, the S3T system we study consists of  mean equation~\eqref{eq:MNSE}
coupled with perturbation covariance equation~\eqref{eq:Lyap1}:
 \begin{subequations}
\label{eq:S3T}
\begin{align}
&\frac{ {d} {\Gamma}}{{d} t} = {G}({\Gamma}) +  {\cal F}( \Cm  )\ ,
  \label{eq:mS3T} \\
& \frac{{d} \Cm }{{d} t} = \Am({U}) \,\Cm+\Cm \,\Am^{\dagger}({ U} ) +  \Qm- \frac{{\rm Tr}(\Qm)}{{\rm Tr}(\Cm)} \Cm\ ,
\label{eq:pS3T}
\end{align}
\end{subequations}
where ${\Gamma} \equiv [{U},{\Psi}]^T$  is the vector of the variables of the  streamwise mean flow, $G(\Gamma)$ expresses the time rate of change of the streamwise  mean flow due to self advection and dissipation, while the term ${\cal F}( \Cm  )$
produces the Reynolds stress forcing of the mean equations from the covariance of the perturbation field, $\Cm$ (see~\eqref{eq:MNSE}). For further details on the formulation see Ref.~\cite{Farrell-Ioannou-2012}.

Results are presented for the minimal Couette flow channel studied by Hamilton, Kim
\& Waleffe~\cite{Hamilton-etal-1995} with streamwise length
$L_x = 1.75 \pi$ and spanwise length $L_z = 1.2 \pi$.
We use $R=600$ (instead of the minimal $R=400$ used in Ref.~\cite{Hamilton-etal-1995}) in order to obtain
turbulence statistics without interruption by relaminarization events.
For examples in which the retained perturbation streamwise wavenumber, $k = 2 \pi / L_x$,  is stochastically excited this is done
using independent compact support wall-normal velocity and
vorticity structures in  $(y, z)$ chosen to inject equal energy into every degree of freedom in the system  as described above.
The resulting spatial forcing covariance, $\Qm$, is spanwise homogeneous and is consistently taken to be the identity matrix. Numerical calculations  employ $N_y=21$ grid points in the wall-normal direction and $N_z=30$ grid points in the spanwise  direction. A study of  S3T turbulence under similar conditions in various channel sizes were reported by Thomas et al.~\cite{Thomas-etal-2015}.

\section{Isolating the linear dynamics of the second order cumulant}

The unforced S3T equations: \begin{subequations} \label{eqn:2RNL1}  \begin{align}
&\frac{ {\df} {\Gamma_a}}{{\df} t} = {G}( {\Gamma_a}) +   {\cal F}( \Cm_a )~,
 \label{eqn:m1}  \\
& \frac{{\df} \Cm_a }{{\df} t} = \Am( {U_a} ) \,\Cm_a+\Cm_a \,\Am^{\dagger}({ U_a} ) \ ,
\label{eqn:p1}
 \end{align}
 \end{subequations}
form  a non-linear dynamical system that self-sustains S3T turbulence
\cite{Farrell-Ioannou-2012,Thomas-etal-2014,Thomas-etal-2015,Bretheim-etal-2015,Farrell-etal-2016-VLSM}.
The quasi-linear structure of this system
allows us to isolate the linear dynamics of the incoherent  component of the turbulence, $\Cm_{b}$:
\begin{equation}
\frac{{\df} \Cm_b }{{\df} t} =\Am ({U}_b) \,\Cm_b+\Cm_b\,\Am^\dagger({ U}_b)\ , \label{eqn:p2}
\end{equation}
 where ${U}_b(y,z,t)$  could be an arbitrary time-dependent mean streamwise velocity  but for our purposes  is taken  to be the solution ${U}_a(y,z,t)$  obtained from a sufficiently long time series of a  self-sustaining turbulence solving~\eqref{eqn:2RNL1}.
With $U_b$ chosen to be identical to the fluctuating mean flow,  $U_a$, of the
self-sustaining S3T turbulent state, the time dependent linear equation~\eqref{eqn:p2} can be verified to have  exactly zero Lyapunov exponent and the covariance,
$\Cm_b$, if randomly initialized can be verified to asymptotically  approach
the rank 1 covariance produced by the structure associated with this zero Lyapunov exponent, which will be referred to as the first Lyapunov vector
(cf.~Appendix~\ref{ap:A}). From the theory of time-dependent linear dynamical systems we  know that as $t \rightarrow \infty$ the covariance can be decomposed into a basis of  time dependent
Lyapunov vectors ordered in average growth rate
by their Lyapunov exponents~\cite{Oseledets-1968,Lorenz-1984, Wolfe-Samelson-2007} (cf.~Appendix~\ref{ap:A}).
This result obtained in the case
of a time dependent linear dynamics
is  analogous  to the more familiar  case of a time independent linear dynamics
in which as $t \rightarrow \infty$ the analogous covariance can
be decomposed into a basis of orthogonal time independent vectors   which, with the exception of the first,
are not identical to the eigenvectors of the associated time independent dynamical operator
but are ordered in growth rate by the  associated dynamical operator's eigenmode
growth rates.
In both the autonomous and non-autonomous case the covariance is exponentially
dominated by the most unstable of these
which has the structure of the most unstable Lyapunov vector and eigenmode respectively.

\begin{figure}
\includegraphics[width =\columnwidth]{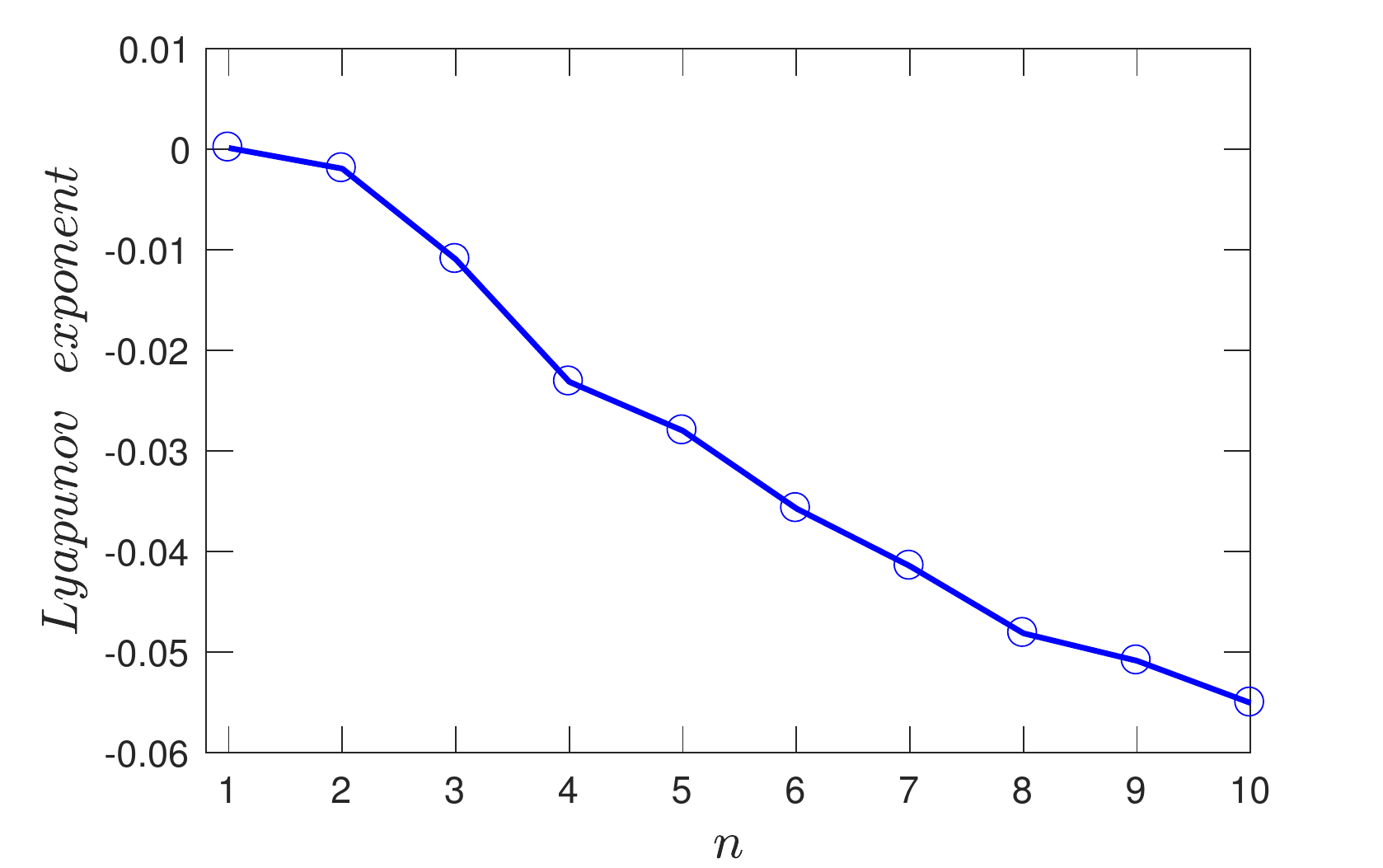}
\caption{\label{fig:lyap_spectrum}  The first ten Lyapunov exponents of the dynamical operator $\Am({U}_a)$. The maximal Lyapunov exponent is zero, consistent with $U_a$ being the consistent time dependent mean streamwise component supporting the turbulent perturbation component of the  combined turbulent state. }
\end{figure}

Consider forcing  the secondary perturbation dynamics~\eqref{eqn:p2} to have the same time dependence as the primary self-sustaining S3T  turbulent system
\eqref{eqn:2RNL1} by setting ${U}_b={U}_a$ in~\eqref{eqn:p2}.  The first
question we address is whether this coupling results
in synchronization of the perturbation fields.
Under forcing by $U$,the first 10 Lyapunov exponents of  a randomly initialized $\Cm_b$, are shown in Fig.~\ref{fig:lyap_spectrum}.
The maximal Lyapunov exponent of $\Cm_b$ assumes the same zero value as that of  $\Cm_a$ and both $\Cm_a$ and $\Cm_b$ assume asymptotically the structure associated with the same corresponding first Lyapunov vector.  However,  $\Cm_a$ and $\Cm_b$
differ in amplitude (to the degree the random  initial state of $\Cm_b$ projects on
the first Lyapunov vector).  Therefore it is required to use  as a synchronization condition convergence of the normalized covariances:
\begin{equation}
 \lim_{t \to \infty} \delta (t) \equiv \lim_{t \to \infty} \left\| \frac{\Cm_a}{\|\Cm_a\|} - \frac{\Cm_b}{\|\Cm_b\|} \right\| = 0\ . \label{eq:delta}
\end{equation}

\begin{figure}
	\includegraphics[width = 0.9\columnwidth]{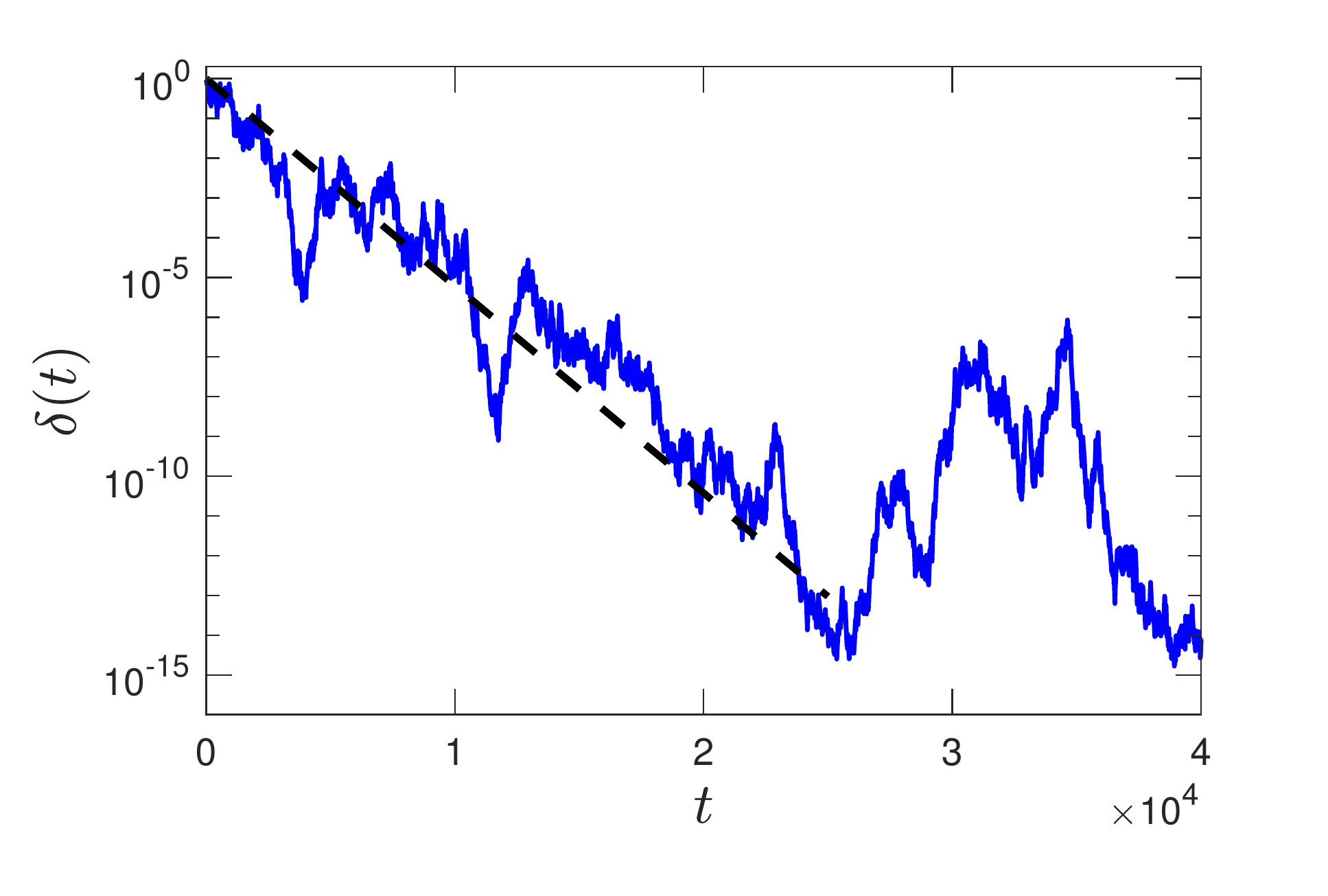}
	\caption{\label{fig:delta} The approach towards synchronization  measured by  $\delta(t)$ occurs at twice the rate of the second Lyapunov exponent of  $\Am(U_a)$, which is indicated with the dashed  line.}
\end{figure}

\begin{figure*}
	\centering
	\includegraphics[width = .6\textwidth]{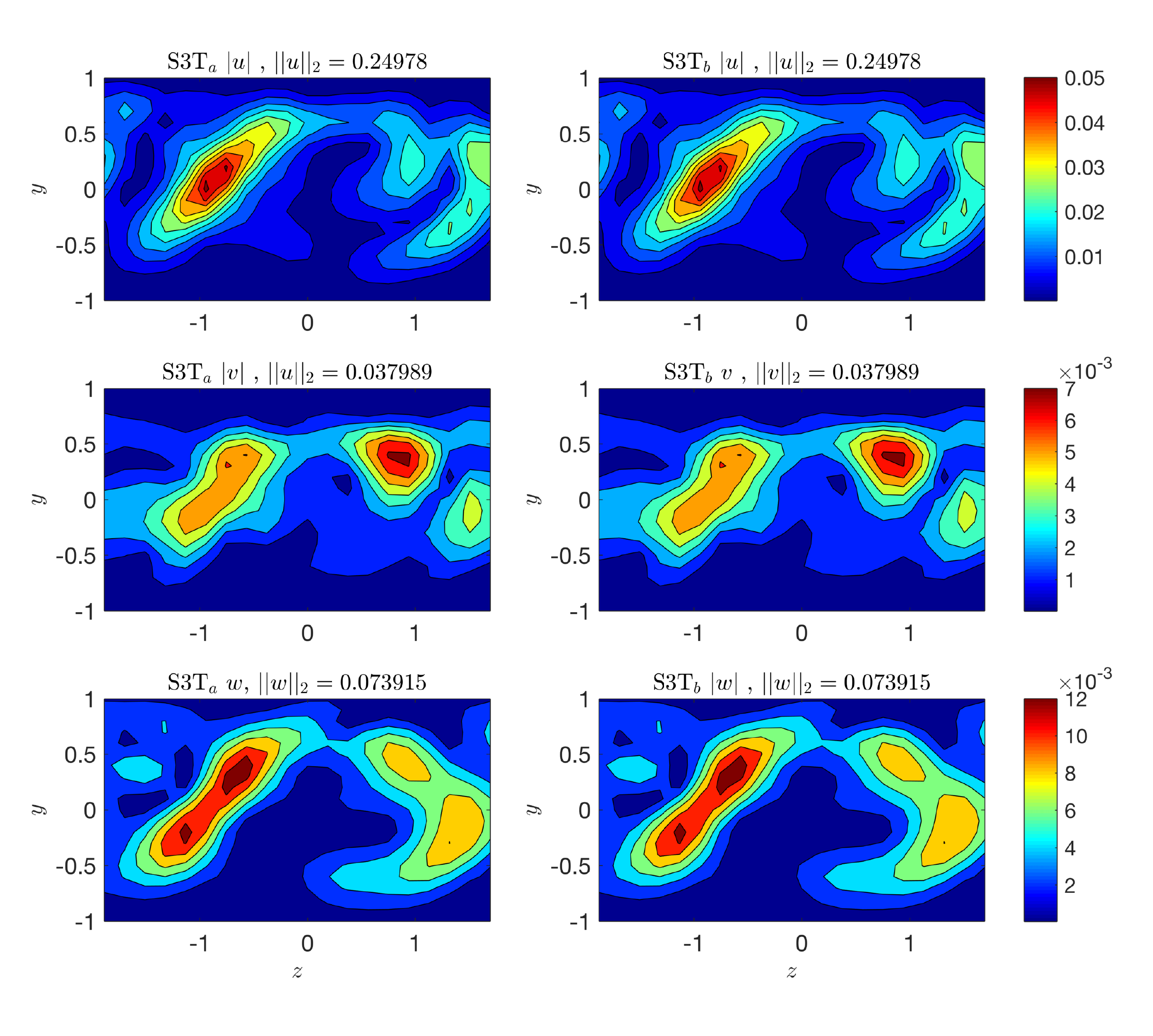}
	\caption{\label{fig:Csync} The three components of perturbation velocity, $u$,$v$, and $w$ (in sequence from top to bottom) of the  first Lyapunov vector (LV$_1$), which is the eigenvector with the largest eigenvalue of the covariance $\Cm_{a}$ of the primary system   (left panels) and similarly for the secondary system with covariance $\Cm_{b}$ (right panels)  initialized  with a different initial condition. The snapshots are  at $t=3\times 10^4$, which is a sufficient time for the asymptotic state to be obtained. The figure demonstrates that in this S3T turbulent state the time varying mean streamwise streak velocity results asymptotically in a unique (to within a streamwise phase) perturbation state to which initial conditions converge. The normalized velocities are represented by contours of their absolute magnitude.}
\end{figure*}

Convergence in this measure proceeds on average  at twice the rate of the decaying second Lyapunov exponent of  $\Am(U_a)$, as shown in  Fig.~\ref{fig:delta}. To within  streamwise phase  the first Lyapunov vector of the primary system, which  is  the top eigenvector of $\Cm_{a}$, is identical to the first Lyapunov vector of the secondary system, which is  the top eigenvector of $\Cm_{b}$, as shown in Fig.~\ref{fig:Csync}.

\begin{figure}
	\includegraphics[width = \columnwidth]{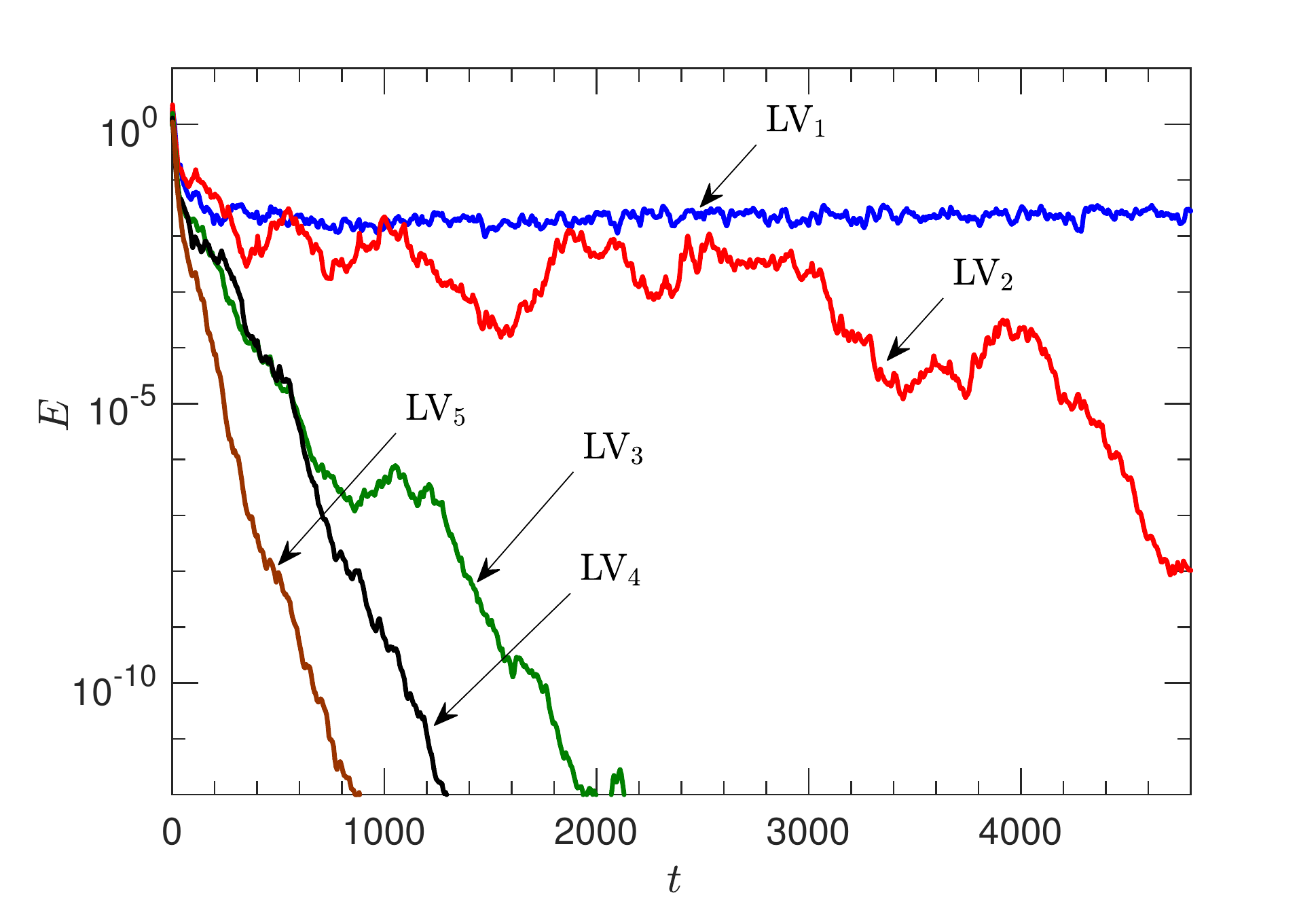}
	\caption{\label{fig:Lyap_decay_rnl}   Evolution of the energy of the first 5 Lyapunov vectors~(LV) of the perturbation covariance dynamics which has been  synchronized with the mean flow of the turbulent S3T system~\eqref{eqn:m1} and initialized white in energy. The streamwise wavenumber of all the Lyapunov vectors is $k= 2 \pi/L_x$  so the individual members of the orthogonal set of Lyapunov vectors differ only in their $y-z$ structure. The maximal Lyapunov exponent associated with  the first Lyapunov vector (LV$_1$) is zero consistent with it constituting  a component of the statistical steady state. Except for LV$_1$, the remaining Lyapunov vectors (LV's)  decay at the rate of their negative Lyapunov exponents. However, the second Lyapunov vector (LV$_2$) has a small negative Lyapunov exponent and exhibits large excursions associated with the time dependence of the dynamical operator.}
\end{figure}

While  both $\Cm_a$ and $\Cm_b$ are with exponential accuracy  rank one, as they are both the covariance produced by the first Lyapunov vector, eigenanalysis of either $\Cm_a$ or $\Cm_b$ reveals the remaining Lyapunov vectors of the linear time dependent system  \eqref{eqn:p2}
which are decaying with time at the rate of their Lyapunov exponents as shown in  Fig.~\ref{fig:lyap_spectrum}.
This decay of the Lyapunov vectors in the order of their (negative) Lyapunov exponents is shown in Fig.~\ref{fig:Lyap_decay_rnl}.
We remark that support of the turbulence by the single top Lyapunov vector is obtained when scattering by the perturbation--perturbation
nonlinearity is ignored ($\Qm=0$ in~\eqref{eq:S3T}). This turbulence provides an opportunity to study the physical mechanisms of
self-sustaining turbulence in maximally simplified form. We will relax the assumption $\Qm=0$ after our initial study of this maximally
simplified self-sustaining state in order to study the effect of perturbation--perturbation nonlinearity on the turbulence and specifically
the role played by the remaining Lyapunov vectors when these are maintained by excitation parameterizing scattering of energy by the
perturbation--perturbation nonlinearity.

In this section we have verified that  the perturbation structure in  S3T
turbulence can be analytically identified with the  first Lyapunov vector of the
time dependent perturbation operator, $\Am(U)$, linearized about the instantaneous streamwise mean flow $U(y,z,t)$.
This result shows that the perturbation variance in S3T turbulence is supported by an identifiable  rank one structure: the top Lyapunov vector of $\Am(U)$.
This perturbation structure has zero Lyapunov exponent and in that sense it
can be understood to be the mode the reduction to neutral stability of which establishes the statistical state of S3T turbulence
corresponding to neutrality of the time dependent streamwise mean velocity.

\section{Energetics of the Lyapunov structures underlying the perturbation component of S3T turbulence}\label{sec:structure}

Having determined the analytical structure of the perturbation field to be that of the top Lyapunov vector of the fluctuating perturbation dynamics we consider
next the mechanism maintaining this structure.  We will show that the perturbation component of the turbulence is not maintained by temporal mode instability but rather by  parametric instability. Parametrically unstable systems are unstable due
to the interaction between the non-normality and the time dependence of the dynamical operator rather than to temporal mode instability of  the operator
at individual instants of time. In fact, instability of the operator is irrelevant to parametric instability as the familiar analysis of parametric instability in the damped pendulum using the Mathieu equation demonstrates.

In order to  study this parametric instability mechanism in detail
the following analysis is performed: at each instant the normalized perturbation
state, $\phi$,  is projected on the ellipsoid the principal axes of which are in the
directions of the eigenvectors of the symmetric matrix $\Am+\Am^\dagger$.
By eigen-decomposition $\Um\Sm\Um^\dagger=\Am+\Am^\dagger$  with
$\Um$  the matrix composed of the instantaneous eigenvectors of $\Am+\Am^\dagger$
arranged in columns and $\Sm$ the diagonal matrix of the corresponding eigenvalues.
The  instantaneous growth rate of perturbation energy is given by
\begin{equation}
	g(t)=\frac{\phi ^\dagger\Um \Sm \Um^\dagger \phi}{ \phi^\dagger \phi }\ .
\end{equation}
Similarly,  we can  calculate  the growth  rate of the perturbation energy that
would be obtained if the eigenmodes were orthogonal with their same eigenvalues by forming the ellipsoid the principal axes of which correspond to the instantaneous growth rate of the eigenmodes of $\Am$ and projecting the normalized state on these eigenmodes. The normalized projections of the perturbation state $\phi$ on this ellipsoid are then the instantaneous equivalent normal growth rates i.e.~the growth rates that would occur if $\Am$ were a normal matrix with these eigenvalues.
The equivalent normal energy growth rate is given by
\begin{equation}
	h(t)=\frac{\phi^\dagger \Em \Dm \Em^{-1}\phi}{\phi^\dagger\Em^{-1 \dagger}  \Em^{-1}\phi}\ ,
\end{equation}
where $\Em$ is the matrix consisting of the instantaneous eigenvectors of $\Am$ arranged in columns and $\Dm$ is the diagonal matrix of twice the associated modal growth rates of the modes.

\begin{figure}
       \includegraphics[width = \columnwidth]{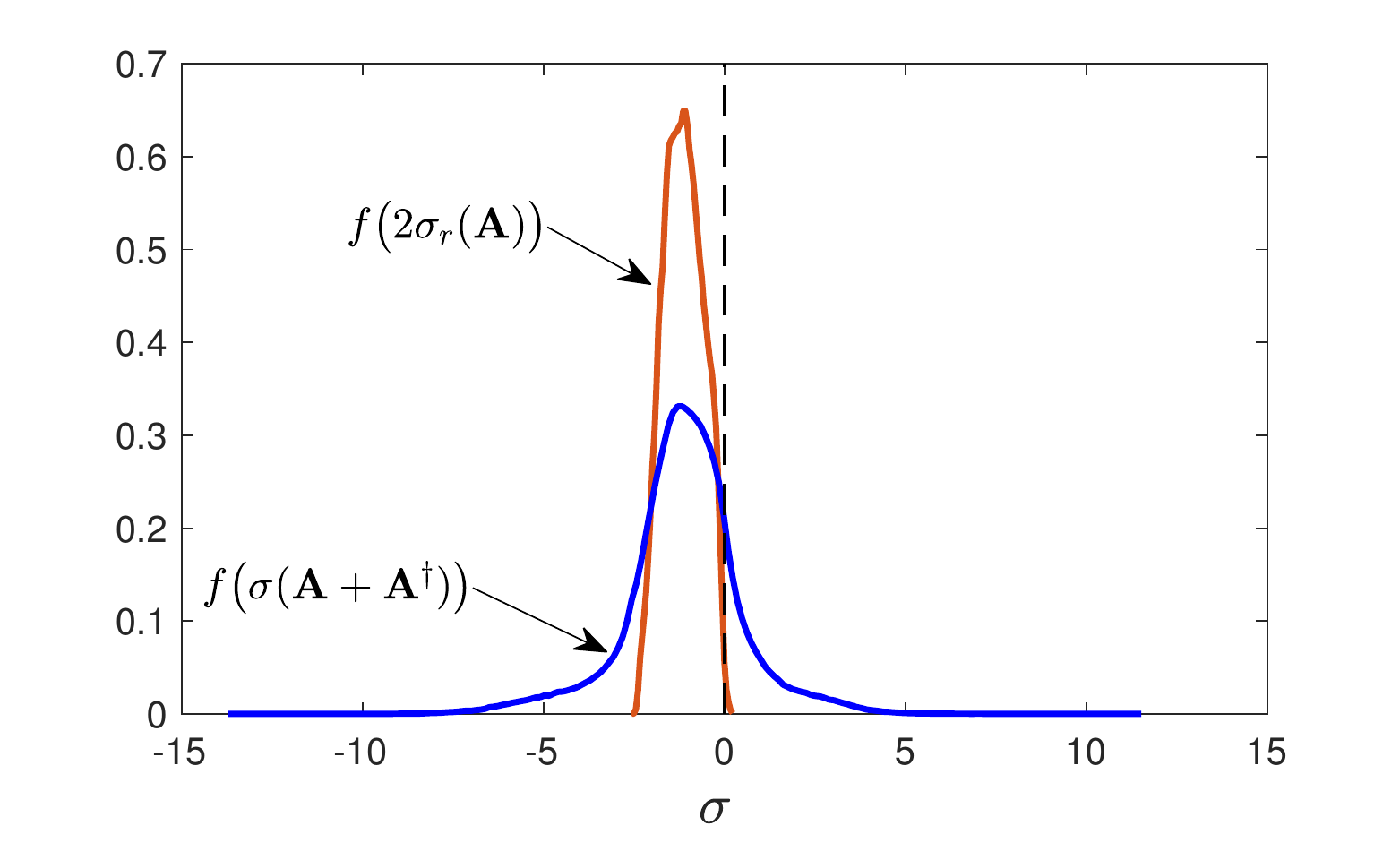}
      \caption{\label{fig:prob_eig} The probability density function of the eigenvalues of  the instantaneous perturbation operators $\Am+\Am^\dagger$ and
      of twice the real part of the instantaneous
      eigenvalues of $\Am$ for mean states that occur over a time period $\tau=5000$.
      The mean growth rate for both cases is $-1.14$ (as the real part of the trace of $\Am$ is equal to the trace of  $(\Am+\Am^\dagger)/2$),
      the standard deviation of $\sigma(\Am+\Am^\dagger)$ is 1.6 and the range in the specific
      simulations is $[-18.2, 16.2]$, while the  standard deviation of
      $2 \sigma_r(\Am)$ is 0.5 and the range is $[-2.9,0.4]$.
      The eigenvalues of $\Am+\Am^\dagger$ correspond to the axes of the energy growth rate ellipsoid.
      The extrema of these possible growth rates exceed that 	of the instantaneous energy eigenfunction growth rates
      as  expected for a non-normal system.   Remarkably,  only very small positive modal growth rates occur suggesting that the system is
      constrained to limit the extent of modal instability.}
        \end{figure}

\begin{figure}
       \includegraphics[width = \columnwidth]{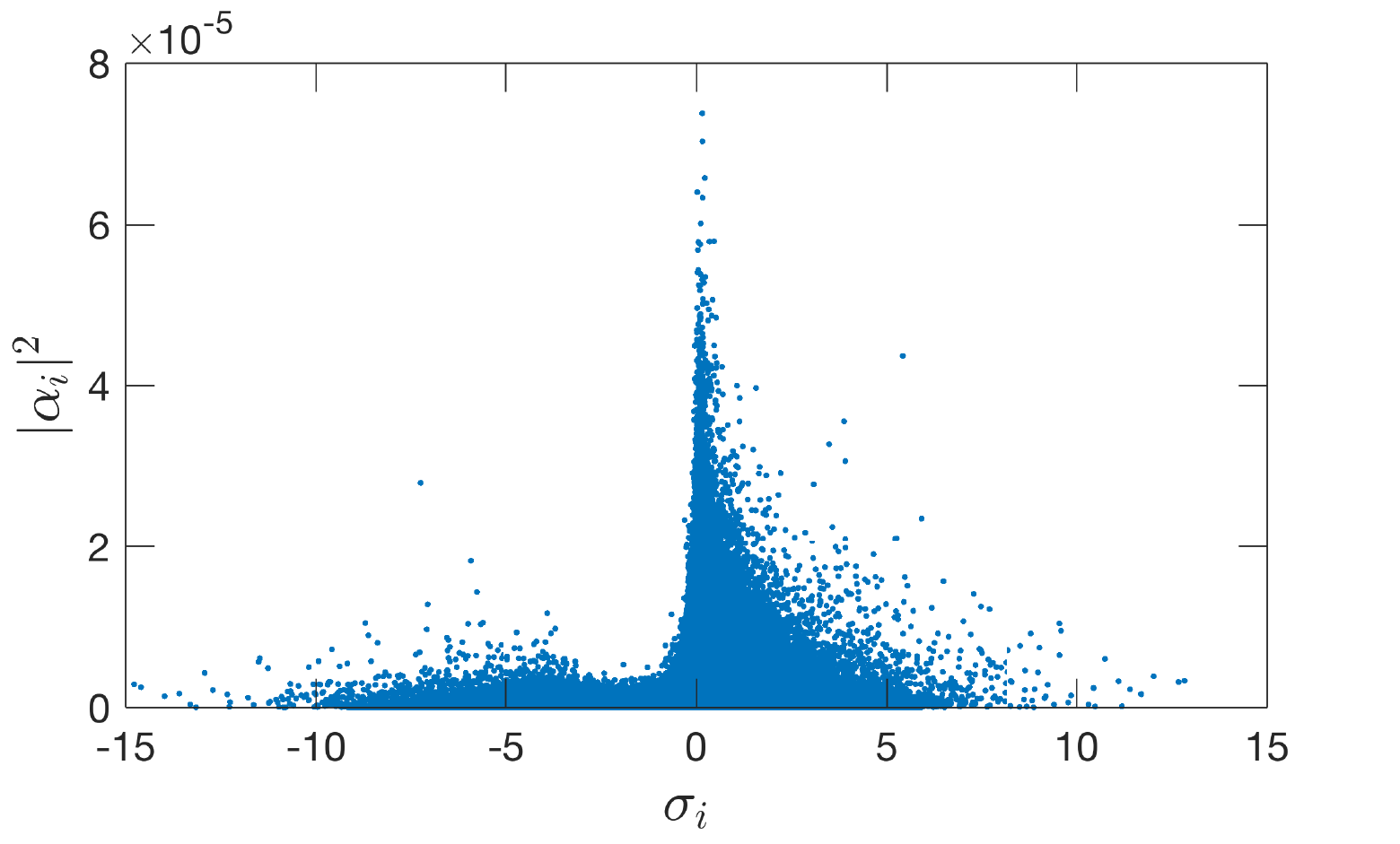}
      \caption{\label{fig:project_ellipsoid}  Instantaneous projections of the normalized state vector on the  axes
       of the ellipsoid of  energy growth rate.  Each point has coordinates $(\sigma_i, |\alpha_i|^2)$  where $\sigma_i$ is the $i$-th eigenvalue of
         $\Am +\Am^\dagger$ and $|\alpha_i|^2$ is the square amplitude of the  projection of the normalized state on this principal axis.  This figure reveals both the magnitude of the projection of the state on the growth rate axes and, by the density of the points, the frequency of the occurrence of each growth rate.}
        \end{figure}

\begin{figure}
       \includegraphics[width = \columnwidth]{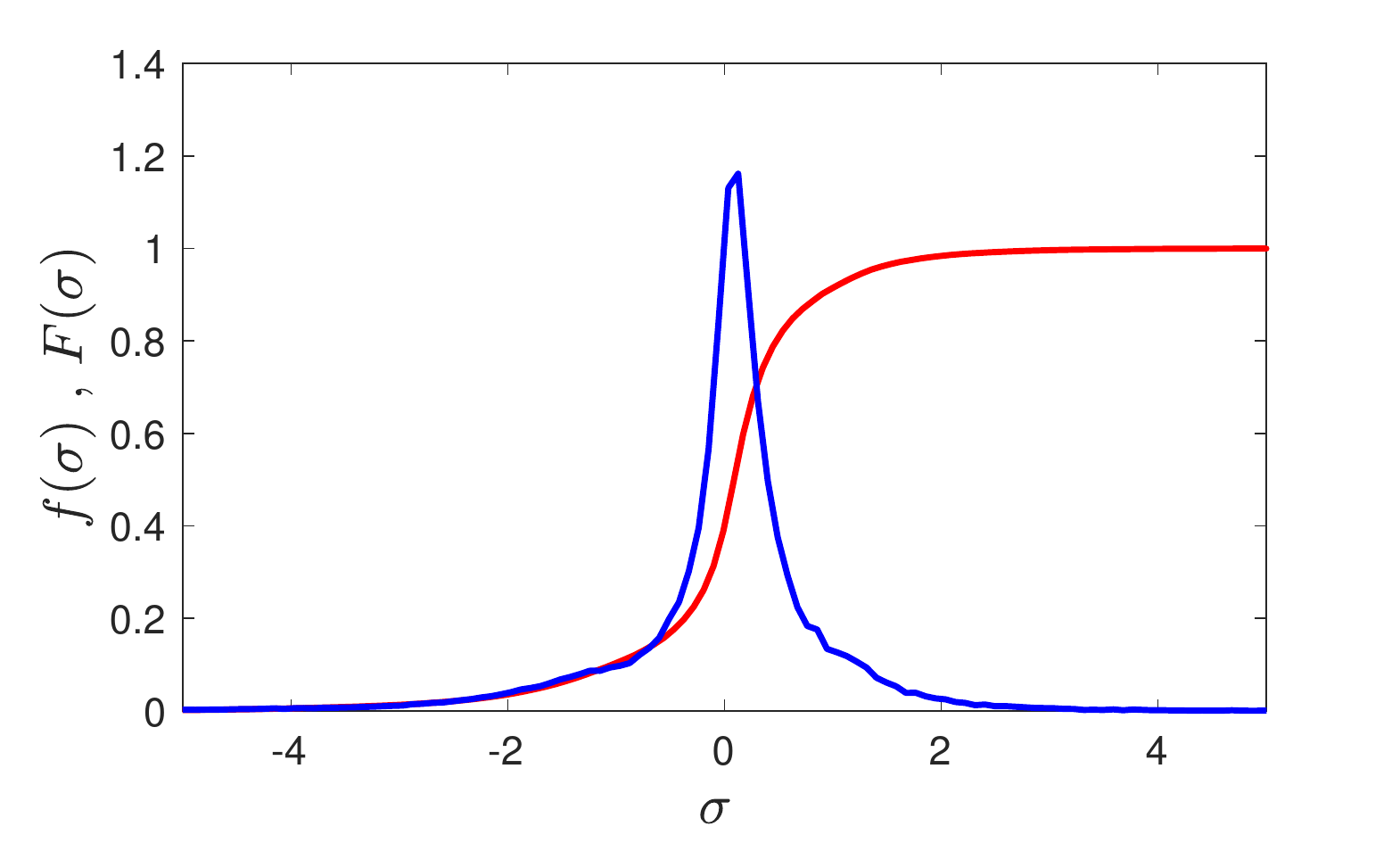}
      \caption{\label{fig:CDF}
      Cumulative distribution function, $F(\sigma)$, of the square projection of the normalized perturbation state consisting of the first Lyapunov vector
      on the  axes
       of the growth rate ellipsoid (red).
       The smooth derivative of the cumulative distribution function, $f(\sigma)$, (shown in blue),  is the probability density function of the perturbation state projections
       $|\alpha(\sigma)|^2$ on the energy growth rate ellipsoid.
}
        \end{figure}

\begin{figure}
       \includegraphics[width = \columnwidth]{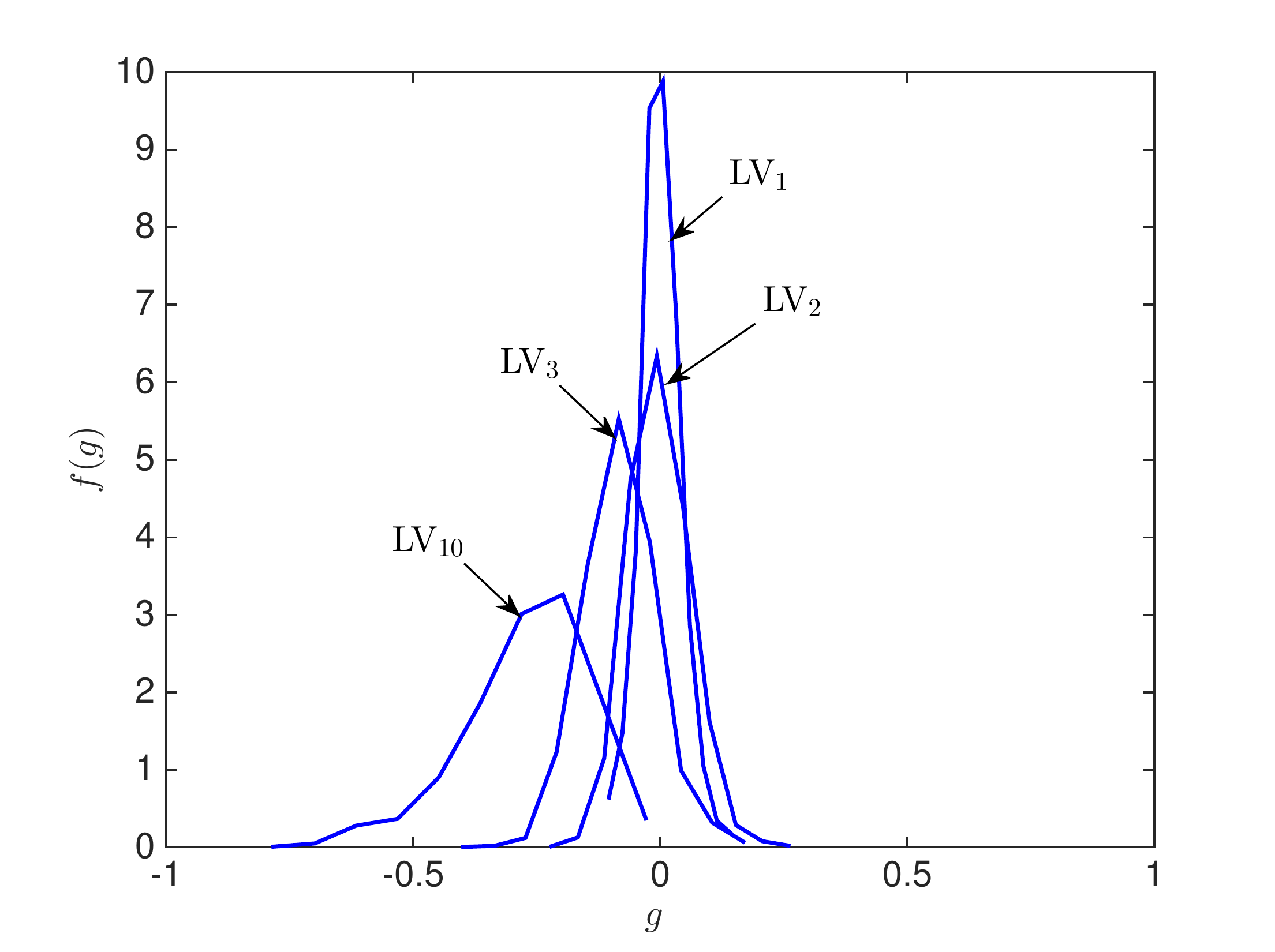}
      \caption{\label{fig:pdf_wall}
         The probability density function of the instantaneous energy growth rates of Lyapunov vectors   LV$_1$, LV$_2$, LV$_3$ and  LV$_{10}$.
      The second in growth rate Lyapunov vector, LV$_2$, is only slightly decaying
      and has a narrowly confined distribution similar to that of the first Lyapunov vector, LV$_1$, while the LV$_3$ and LV$_{10}$ decay strongly and sample a wider range of growth rates. }
        \end{figure}

\begin{figure}
       \includegraphics[width = \columnwidth]{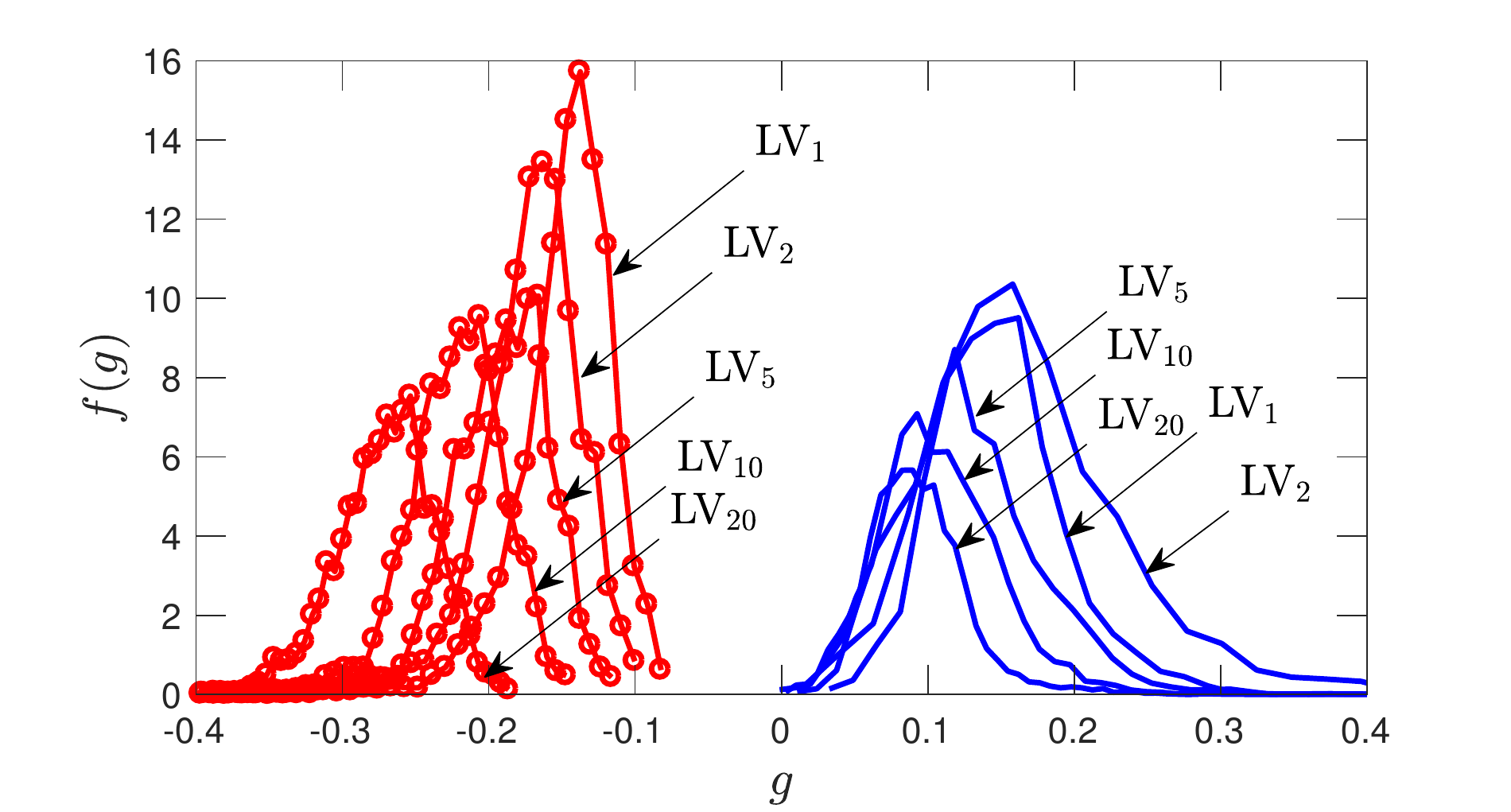}
      \caption{\label{fig:glyap} Partition of the probability density function of the
      instantaneous energy growth rates of Lyapunov vectors LV$_1$, LV$_2$, LV$_5$, LV$_{10}$ and LV$_{20}$ arising from energy transfer to these vectors from interaction with the streamwise mean flow (curves on the right) and dissipation (curves on the left). The probability density function of the instantaneous growth rates of each Lyapunov vector, shown in Fig.~\ref{fig:pdf_wall}, is the sum of these contributions for each LV. A substantial number of higher order Lyapunov vectors robustly extract energy from the mean. In fact, half the LV's extract some energy directly from the mean flow.}
              \end{figure}

\begin{figure}
       \includegraphics[width = \columnwidth]{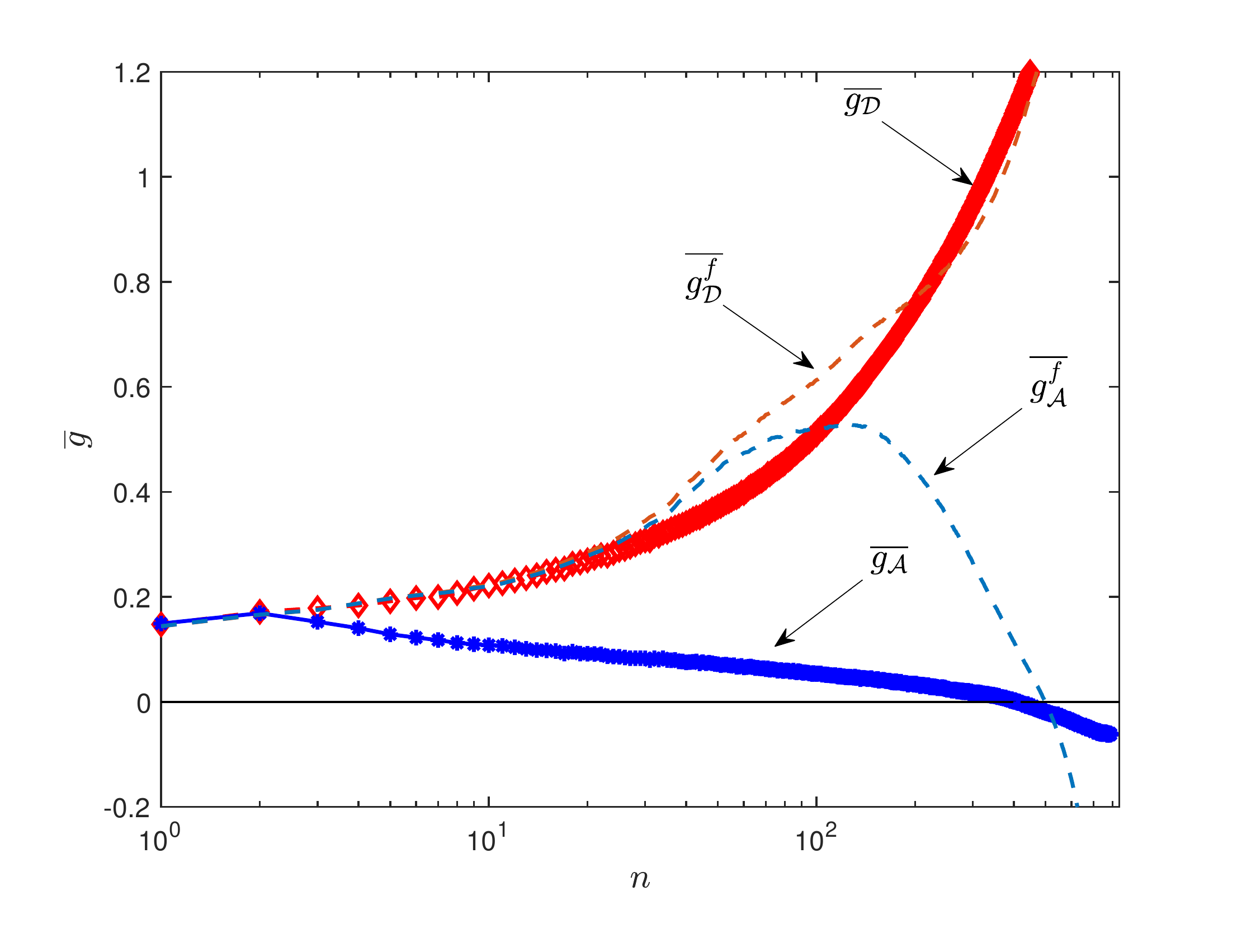}
      \caption{\label{fig:glyap_all}Mean contribution to the energy growth rate  of the Lyapunov vectors due to energy
      transfer from  the mean flow,
      $\overline{g_{\cal A}}$ as a function of Lyapunov vector index. Also shown is the magnitude of the
       mean decay rate of the Lyapunov vectors due to diffusion,  $|\overline{g_{\cal D}}|$. The Lyapunov exponent of
       each Lyapunov vector is the difference between these two curves. The dashed curve,  $\overline{g_{\cal A}^f}$, shows the average
       energy     transfer rate from the
       mean to each of the orthogonal eigenfunctions of the covariance $\Cm$ under external excitation (the SFLV's of the
       stochastically maintained perturbation field)
       and the dashed curve, $|\overline{g_{\cal D}^f}|$,  the average decay rate of each of the SFLV's.
       This figure shows that a substantial subset   of SFLV's have neutral energetics and therefore are
       maintained at finite
       amplitude by transfer of energy from the mean flow induced by the energy neutral parameterization of  perturbation--perturbation
       nonlinearity  (the asymptotic rank of $\Cm$ is about 50).}
        \end{figure}

The probability density function  of the eigenvalues of $\Am+\Am^\dagger$, which
correspond to the axes of the instantaneous growth rate ellipsoid, and the probability density function of
twice the real part of the  eigenvalues of $\Am$, which correspond to the
axes of the modal growth rate ellipsoid, are shown for a self-sustaining turbulent state in
the example system over a time interval $\tau=5000$ in Fig.~\ref{fig:prob_eig}. The instantaneous growth rate  of the perturbation state is determined by its projection on the instantaneous growth rate ellipsoid.  This projection varies in time due to both the time dependence of the state vector and the time dependence of the growth rate ellipsoid. The distributions of the resulting projections  for a turbulent simulation over a time period $\tau=5000$ is shown in Fig.~\ref{fig:project_ellipsoid}. This figure contains information on both the extent of the growth rates sampled by the state vector as well as the frequency with which these values are sampled. The state vector fails to explore the extremities of the growth rate ellipsoid with most projections being confined around zero growth rate. The information in Fig.~\ref{fig:project_ellipsoid} is summarized by the cumulative distribution function  of the square projections of the state on the  principal axes of the growth rate ellipsoid shown in Fig.~\ref{fig:CDF}. This cumulative distribution function  is obtained from Fig.~\ref{fig:project_ellipsoid} by forming
\begin{equation}
	F(\sigma)  = \dfrac{\int_{-\infty}^{\sigma}\delta(\sigma'-\sigma_i)  |\alpha_i|^2 \,\df\sigma' }{\int_{-\infty}^\infty \delta(\sigma'-\sigma_i)  |\alpha_i|^2\,\df\sigma'}\ ,
\end{equation}
in which each of the points in Fig.~\ref{fig:project_ellipsoid} is a sample $(\sigma_i,|\alpha_i|^2)$.
The smooth derivative of the cumulative distribution function,
$f(\sigma)=dF(\sigma)/d\sigma$,
also shown in Fig.~\ref{fig:CDF},  is the probability density
function of the perturbation state projections, $|\alpha(\sigma)|^2$, on the
energy growth rate, $\sigma$. Despite the wide distribution of available growth
rates (cf.~Fig.~\ref{fig:prob_eig}) the self-sustained state projects  on growth rates
narrowly centered around zero with values primarily in the interval
$[-1,1]$. The mean growth rate of the state,
\begin{equation}
	\lambda = \int_{-\infty}^\infty \sigma f(\sigma)\,\df\sigma\ ,
\end{equation}
vanishes consistent with the perturbation being a component of the statistically stable turbulent state trajectory, i.e.~the state trajectory, corresponding to the first Lyapunov vector (LV$_1$), has been adjusted, together with and by mutual interaction with the mean flow, to have zero Lyapunov exponent. An equivalent diagnostic, the growth rate probability density function, is more easily obtained directly from the time series of the growth rates of the individual Lyapunov vectors (LV's).
This probability density function is shown for LV$_1$, and also for the decaying second, third
and tenth Lyapunov vectors, LV$_2$, LV$_3$, and LV$_{10}$,  in Fig.~\ref{fig:pdf_wall} in
which it can be seen that, although the probability density function of LV$_1$ peaks at positive growth rates
it has a small negative skew and  its mean growth rate is zero and although LV$_2$ has a negative Lyapunov exponent, it is similar to LV$_1$ in its energetics.
Energetics of the Lyapunov vectors can be more closely analyzed by separating the operator
of the linear perturbation dynamics into dynamical  and dissipation components by partitioning it as
\begin{equation}
 \Am(U) \equiv  {\cal A}(U)+\cal{D}\ .\label{partition}
\end{equation}
In~\eqref{partition} ${\cal A}(U)$ is the part of the matrix $\Am(U)$ that depends on $U$ and its spatial derivatives and represents dynamic interaction of the perturbation field with the streamwise mean flow, including the transfer of energy between mean and perturbations, and $\cal D$ is the part of the
matrix $\Am$  associated with  viscous damping. The terms involved in this separation can be identified in the pre-transformed perturbation equations~\eqref{eq:PNSE}. With this splitting the rate of  transfer of mean flow energy to  a given perturbation, $g_{\cal A}$,  is given by the normalized projection of that perturbation on the Hermitian matrix $\cal{A} + \cal{A}^\dagger$. Similarly, the rate of dissipation, $g_{\cal D}$,  is given by its normalized projection on $\cal{D}^\dagger + \cal{D}$. The probability density function of $g_{\cal A}$  for a selection of Lyapunov vectors and the probability density function of the corresponding decay rates due to dissipation, $g_{\cal D}$, shown in Fig.~\ref{fig:glyap}, reveals that the asymptotic decay of Lyapunov vectors of order 2 and higher is due to enhanced dissipation  rather than to inability to gain energy from the mean flow.
In fact, mean flow energy is transferred to LV2 at a greater rate than it is to LV1  as seen in Fig.~\ref{fig:glyap}.
In Fig.~\ref{fig:glyap_all} is shown the time-mean growth rate of these Lyapunov vectors due to energy transfer from the mean flow, $\overline{g_{\cal A}}$, and the corresponding magnitude  of their mean energy decay rate due to dissipation, $|\overline{ g_{\cal D}}|$ (overbar denotes time average).
These average growth  and decay rates determine the Lyapunov exponent of the corresponding Lyapunov vector, which is given by
 $\overline{ g_{\cal A}} -|\overline { g_{\cal D} }|$. It is interesting to note that although  all Lyapunov vectors, except LV$_1$, are decaying, exactly half of the Lyapunov vectors receive energy from the mean. This property is a corollary of the ``time-reversal symmetry" of the inviscid perturbation dynamics that is
 governed by  $\cal A$. This symmetry of inviscid dynamics is the
 expression of the invariance  of the perturbation evolution equations~\eqref{eq:PNSE} (in the absence of dissipation or excitation) to the transformation $t \to -t$ and $x \to -x$. This symmetry implies that if $\phi(x,t)$ is a  solution of  the inviscid equations~\eqref{eq:PNSE} without excitation, by necessity  $\phi(-x,-t)$ is also a solution of~\eqref{eq:PNSE} and hence if $\phi(x,t)$
 grows at the rate $\lambda$, $\phi(-x,-t)$ grows at  the same rate and by reversing time, $\phi(-x,t)$  decays at rate $-\lambda$.
 This implies  that if $\phi$ is a Lyapunov vector of ${\cal A}(U)$ with Lyapunov exponent $\lambda$ then $\phi^*$ is also a Lyapunov vector
 with Lyapunov exponent $-\lambda$.
The fact that fully half the perturbation structures extract energy from the fluctuating mean flow has important consequences for the maintenance of perturbation variance in turbulence as it implies a mechanism for direct non-local in scale transfer of energy from the externally forced large scale mean flow to the small scale perturbation components of  the turbulence. This fact invites the conjecture that the mechanism maintaining the incoherent component of perturbation energy in S3T turbulence when parameterization of the third cumulant is restored in the S3T  dynamics is parametric interaction with the mean flow and that the structure of the turbulent perturbations is primarily determined by the structure of the LV's. We will restore  stochastic parameterization of the third cumulant to examine this conjecture further in the next section.

Another implication of the parametric dynamics maintaining the incoherent  perturbation component
is that  the decaying Lyapunov vectors undergo large excursions in energy (cf.~Fig.~\ref{fig:Lyap_decay_rnl}). Such large excursions are characteristic of  the energetics of stochastic dynamical systems with multiplicative noise~\cite{Farrell-Ioannou-1999, Farrell-Ioannou-2002-optimal, Farrell-Ioannou-2002-perturbation-II}. When a parameterization for the excitation of Lyapunov vectors by perturbation--perturbation nonlinearity is included we expect that a broad spectrum of incoherent perturbations will be supported by interaction with the mean flow. Moreover, these structures undergo large excursions and these excursions have important implications for the dynamics of S3T turbulence. We will further explore these matters but first we wish to examine the dynamics of the parametric mechanism in more detail.

  \begin{figure*}
       \includegraphics[width = .6\textwidth]{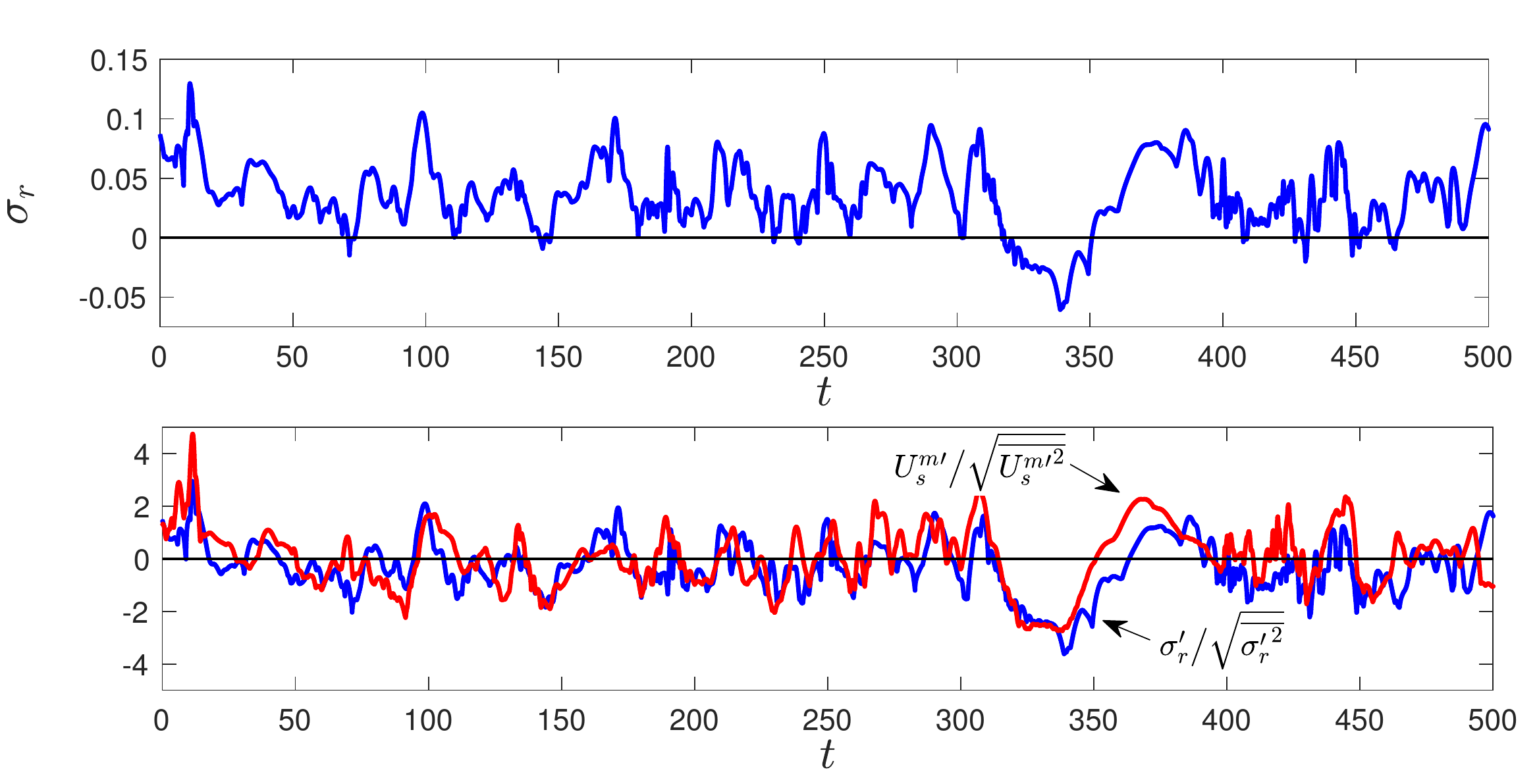}
      \caption{Panel~(a):~The maximum modal growth rate of $\Am(U)$  over a typical interval in the example system. The mean of the maximum growth rate  over the entire time series is $0.045$ and the range is $[-0.06,0.135]$.
       Panel~(b):~Corresponding time series of the normalized fluctuations of the  maximum growth rate, $\sigma_r' = \sigma_r - \overline{ \sigma_r}$, and  of the normalized fluctuations of the maximum streak amplitude  $U_s^m \equiv \max(U_s(y,z) )-\min(U_s(y,z))$. The streak amplitude and the maximum instability growth rate
       are correlated  with correlation coefficient $r=0.4$ consistent with inflectional instability of the streak. However, while the state is tightly constrained to exhibit small but consistent  instantaneous modal instability, this instability  is not responsible for sustaining the perturbation variance.}
      \label{fig:kcimax}
        \end{figure*}

      \begin{figure}
       \includegraphics[width = \columnwidth]{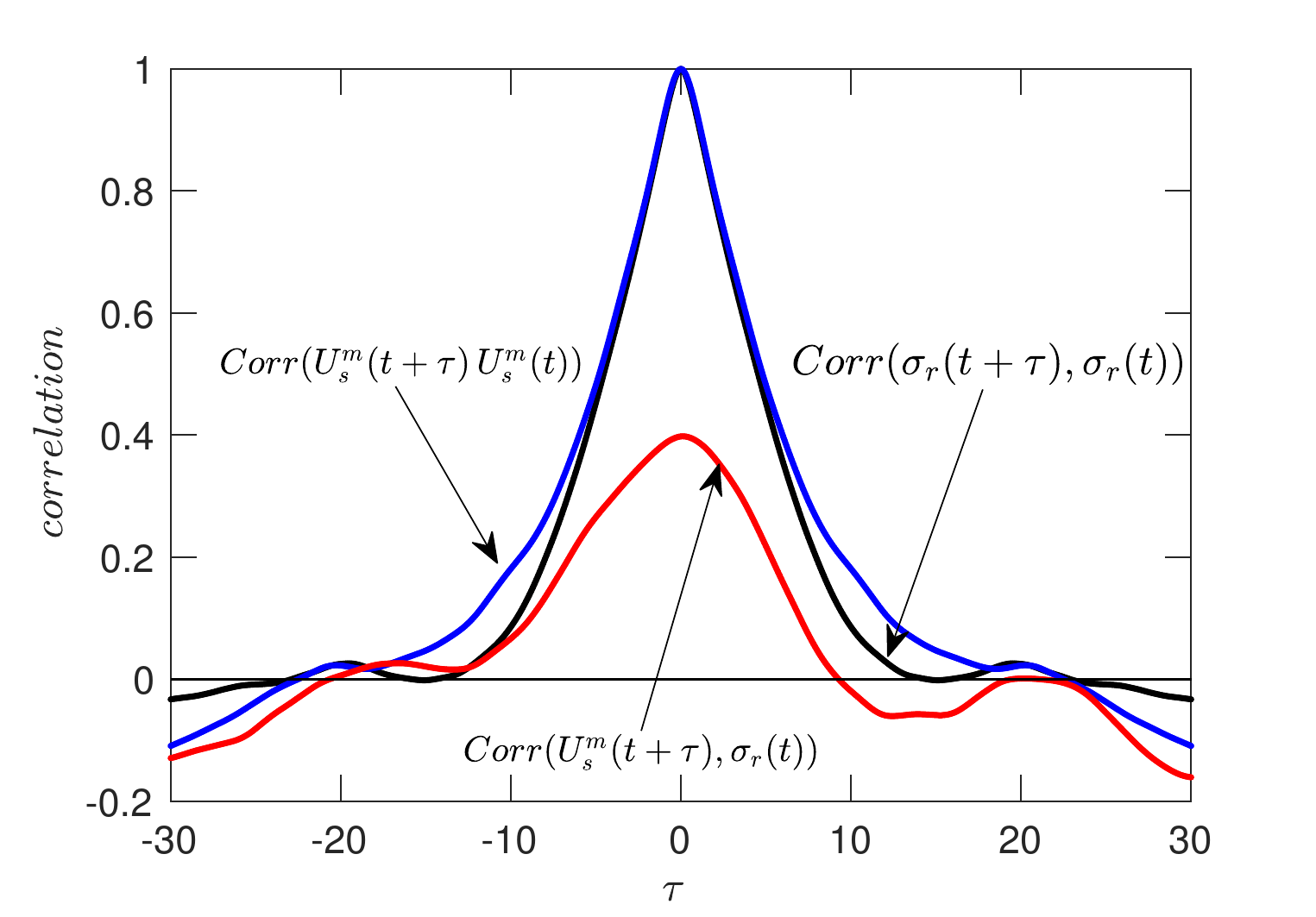}
      \caption{ Autocorrelation of  streak amplitude,   $U_s^m \equiv \max(U_s(y,z) )-\min(U_s(y,z))$, and maximum modal growth rate, $\sigma_r$.  Also shown is the cross correlation of these quantities.  The streak and modal instability are highly correlated as expected for inflectional mode instability but both of these quantities decorrelate in approximately ten time units which is too short a time for the modal instability to emerge given its typical time scale for growth of approximately 20 time units  (cf.~Fig.~\ref{fig:kcimax}).     \label{fig:cross_kcimax} }
        \end{figure}

\begin{figure}
\includegraphics[width = \columnwidth]{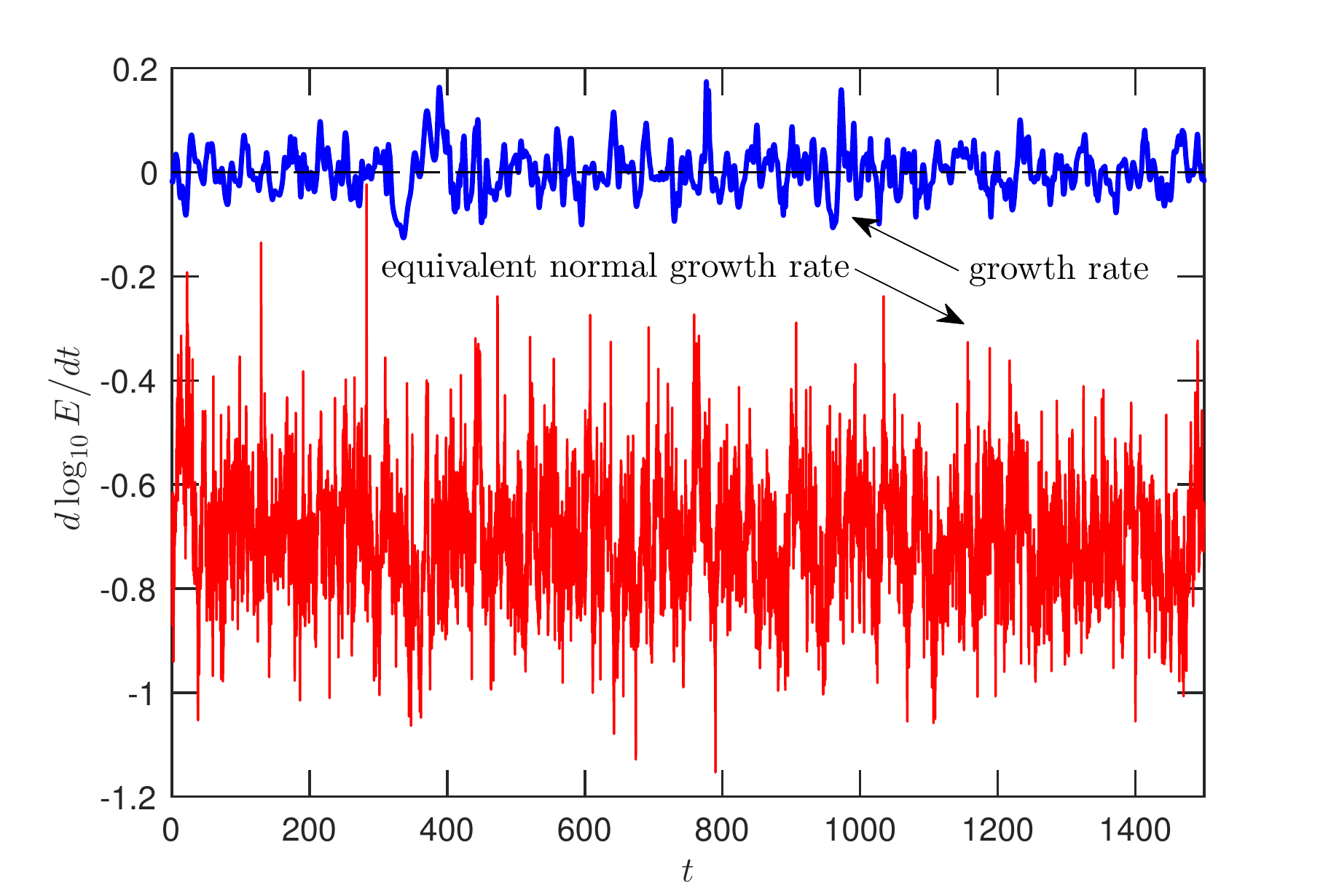}
\caption{\label{fig:gr_ngr} Time series of the energy growth rate of the
perturbation field (blue). This growth rate is equal to the projection of the normalized state on $\Am+\Am^{\dagger}$ and its mean value is the Lyapunov exponent
of LV$_1$, which  is zero. Energy growth rate that would occur if the state were projected on the eigenvectors of instantaneous operator $\Am$ and each advanced at the rate of the corresponding real part of the eigenvalue of $\Am$ is also shown (red). The mean value of this equivalent normal growth rate is $-0.7$. We conclude that while the instantaneous mean states are often modally unstable (cf.~Fig.~\ref{fig:kcimax}) the perturbation state does not project sufficiently on the instabilities to account for its growth. This result demonstrates that the perturbation field is sustained by the parametric non-normal growth process rather than by modal instability.}
\end{figure}

\begin{figure}
\includegraphics[width = \columnwidth]{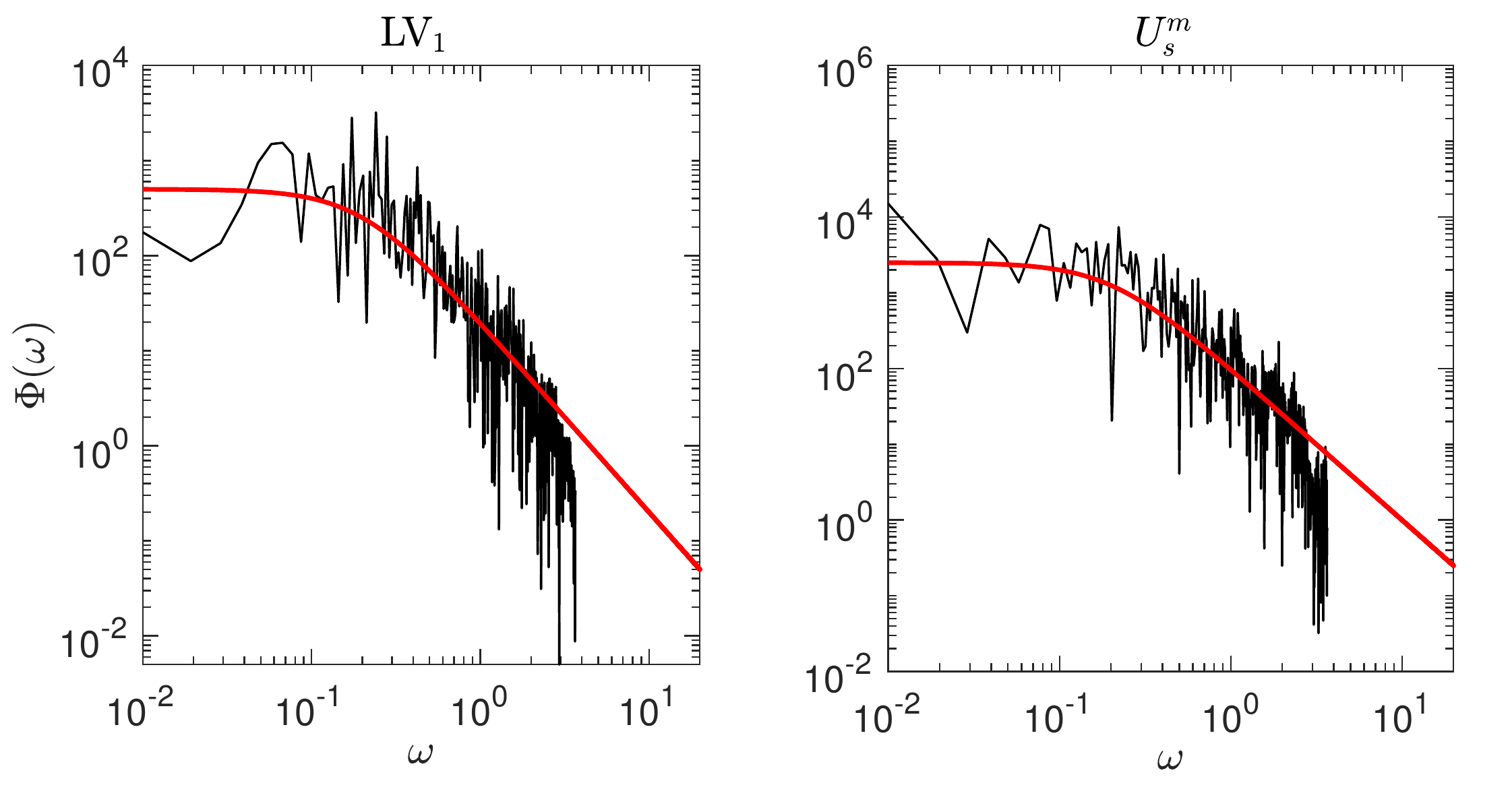}
\caption{\label{fig:spec_den_lv1} Spectral density of the energy growth rate of the first Lyapunov vector (LV$_1$) and of the maximum streak velocity time series and their fit to the Lorentzian $625 \tau^{-2} / ((\omega \tau)^2 +1)$  and $100 U_{s}^m\tau^{2} / ((\omega \tau)^2 +1)$ with $\tau=5.0$, respectively.   This graph shows that the instantaneous growth rate of the perturbations are  well approximated by a red noise process and that the streak fluctuations follow the same red noise process.}
\end{figure}

\begin{figure}
\includegraphics[width = \columnwidth]{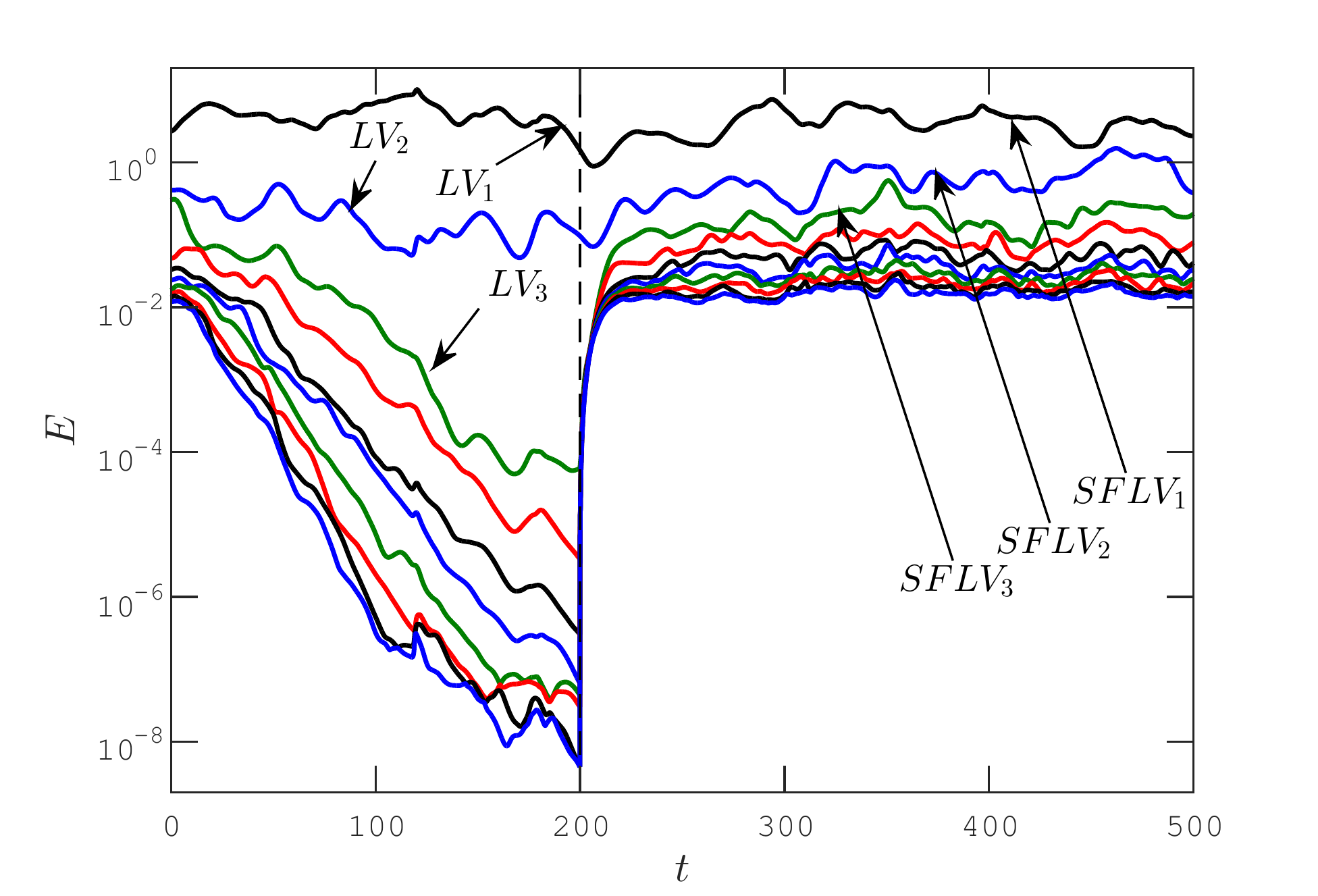}
\caption{\label{fig:sigma0f}   Evolution of the energy of the first 10 eigenfunctions of the covariance $\Cm$ with the highest energy. For $t<200$ the covariance dynamics evolves without stochastic excitation and the structures shown are the first 10 Lyapunov vectors of $\Am(U)$, of which only LV$_1$ would be sustained while the other Lyapunov vectors (even LV$_2$) would decay to zero. A stochastic excitation that imparts no mean energy to the perturbations is introduced at $t=200$. Despite the zero energy input by this parameterization, the non-normality of the time dependent $\Am(U)$ sustains a perturbation covariance of rank approximately 50 supported by SFLV structures close to the corresponding Lyapunov vectors as shown in Fig.~\ref{fig:lv_sflv} and
Fig.~\ref{fig:e_lv_sflv}.}
\end{figure}

\begin{figure}
\includegraphics[width = \columnwidth]{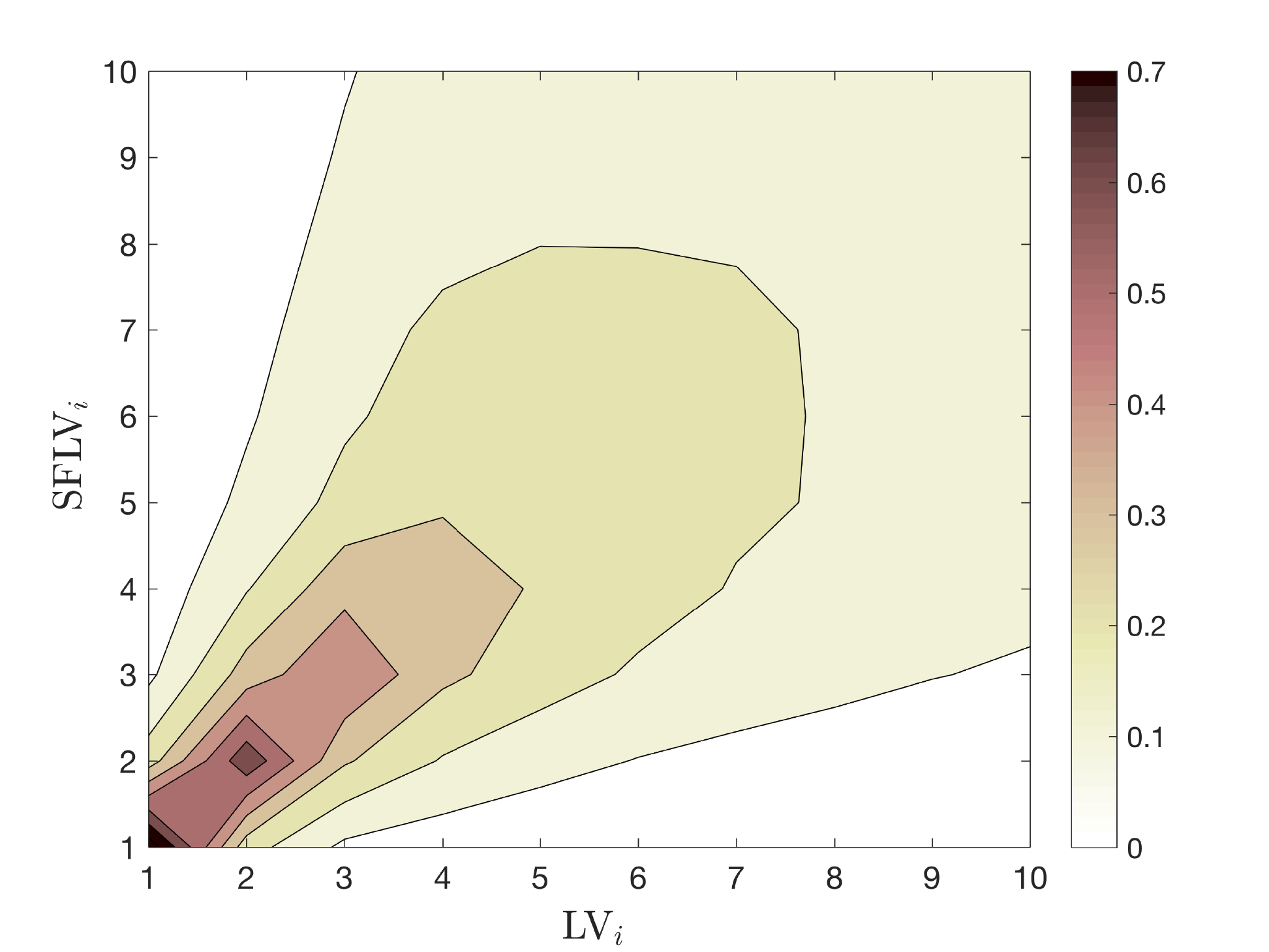}
\caption{\label{fig:lv_sflv} Contour plot of the average absolute value of the inner product, $|{\rm SFLV}_i^\dagger\,{\rm LV}_j |$ with $i,j=1,\dots,10$, between the top 10 SFLV's (the normalized eigenvectors of the $\Cm$ under stochastic excitations) and the first 10 LV's  of $\Am(U)$ with the same streamwise mean flow, $U$. This figure shows that the SFLV's with substantial energy are correlated in structure with  the top LV's. The level of excitation is such that if it were imposed on the time mean flow it would support perturbation energy $1\%$ of the mean energy of the unperturbed Couette flow.}
\end{figure}

\begin{figure}
\includegraphics[width = \columnwidth]{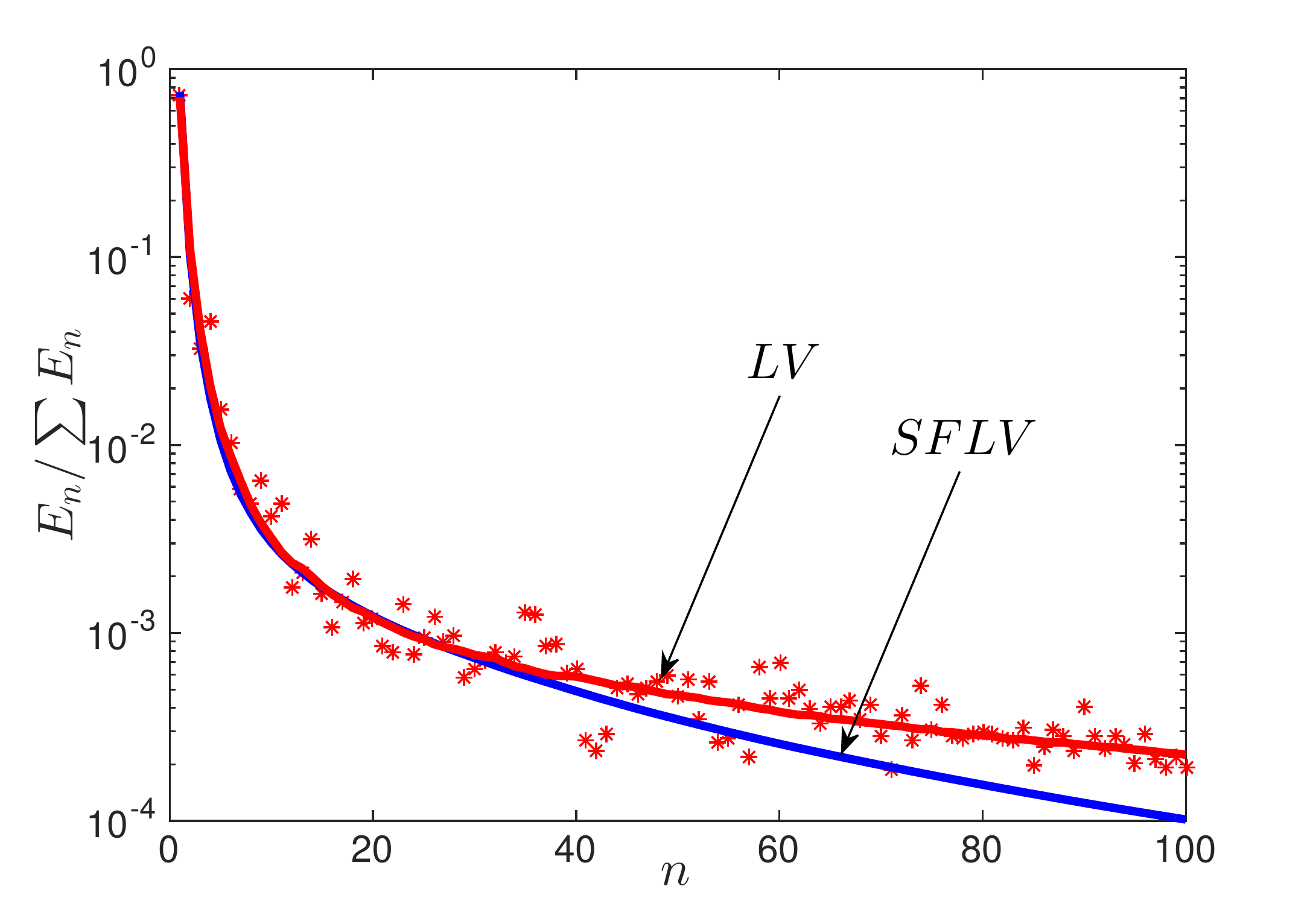}
\caption{\label{fig:e_lv_sflv}{Time averaged fraction of the perturbation energy
accounted by the LV's and SFLV's. The SFLV's are the eigenfunctions of the  velocity covariance, $\Cm$ (the  SFLV set provides the proper  orthogonal decomposition of the flow).
The dots indicate the energy accounted for  by
the LV's at a specific time instant.   This graph shows that
most of the energy is spanned by a  few LV's consistent with the subspaces spanned incrementally by the
LV's being very close to those of the SFLV's (cf. Fig. \ref{fig:lv_sflv}).
The level of stochastic excitation is chosen to maintain perturbation variance typical of turbulence in shear flow i.e. so
that if it were imposed on the time mean flow it would support perturbation
energy $1\%$ of the mean energy of the unperturbed Couette flow}.}
\end{figure}

The linear perturbation dynamics is strongly time dependent with the dynamical operator $\Am$ and associated growth rate ellipsoid being modified continuously in time by fluctuations in the mean streamwise velocity.  These fluctuations cause the dynamics of the interaction between the mean flow and the perturbation field to exhibit large excursions in growth rate on short time scales. In contrast, as shown in Fig.~\ref{fig:prob_eig}, the system adjusts to support only weak instantaneous inflectional instability. A time series of the maximum instantaneous modal
growth rate, $\sigma_r$, is shown in  Fig.~\ref{fig:kcimax}. The mean of the maximum growth rate is only $0.045$ and the growth rate varies rapidly. Also shown is the time series of the normalized fluctuations of the maximum streak amplitude  $U_s^m \equiv \max(U_s(y,z))-\min(U_s(y,z))$ together with normalized fluctuations of the maximum growth rate,  $\sigma_r' = \sigma_r - \langle \sigma_r \rangle$.
The streak amplitude and the maximum instability growth rate are substantially correlated consistent with inflectional instability of the streak. However, while the state dynamics adjusts to consistently exhibit a small and rapidly varying instantaneous modal instability, this instability is not itself responsible for sustaining the perturbation variance. Shown in Fig.~\ref{fig:cross_kcimax} is the autocorrelation and crosscorrelation of  streak amplitude and maximum modal growth rate. The streak and modal instability are correlated with essentially zero lag, as expected for inflectional mode instability,  but both of these quantities decorrelate in approximately ten time units which is inconsistent with emergence of the modal instability which has an approximate e-fold time of 20 (cf.~Fig.~\ref{fig:kcimax}).  That modal growth does not account for maintenance of the perturbation variance is confirmed in Fig.~\ref{fig:gr_ngr} in which is shown a time series of the  energy growth rate of the perturbation field together with the energy growth rate  that would occur if the state were projected on the eigenvectors of the instantaneous operator, $\Am$, with each advanced at the rate of the corresponding real part of the eigenvalue of $\Am$. Although it can be seen from Fig.~\ref{fig:kcimax}  that the instantaneous mean flow is nearly always weakly modally unstable, it can be seen
from Fig.~\ref{fig:gr_ngr} that the contribution to maintaining the Lyapunov vector of the perturbation trajectory arising from modal growth is almost always negative  with the time mean value of this equivalent normal contribution to the energy of the perturbation field being decay at rate  $-0.7$.  In contrast,  the non-normal parametric mechanism produces robust excursions of both positive and negative growth rate due to  rapidly varying projections of the perturbation state on the also rapidly varying directions of growth associated with variation of the streak (cf.~Fig.~\ref{fig:cross_kcimax}).
 Because the maximum Lyapunov exponent is zero and  equal to the time integral of the instantaneous growth rates, these  positive and negative
 contributions average of necessity to zero. The Lyapunov exponent is zero because the associated Lyapunov vector is continuously
 adjusted through interaction with the streak to  have a statistically steady amplitude consistent with its being  a component of the system's
 statistically stable trajectory (cf.~Fig.~\ref{fig:gr_ngr}).
 Parametric growth is a general attribute of
 dynamical systems with stochastically fluctuating dynamical operators which are of necessity non-normal with measure zero exception.  This statistically sustained growth
arises from the concatenation of non-normal growth events which dominate over decay events due to the convexity of the exponential propagator of the dynamics over the time scale of the operator fluctuation~\cite{Farrell-Ioannou-1996b, Farrell-Ioannou-1999}. A characteristic property of
stochastic parametric growth is the requirement for the parametric variation of the system to occur on intermediate time scales. This is because the convexity of the exponential vanishes at short dynamical operator fluctuation time scales and the transient perturbation growth vanishes at long time scales. Fluctuations of the streak and the fluctuation in the growth rate of LV$_1$ fit a red noise process as shown in Fig.~\ref{fig:spec_den_lv1}. Consistent with the stochastic parametric growth mechanism, the correlation time of this red noise process, $\tau=5.0$, occurs on an intermediate time scale. Moreover, this time scale is short compared to the modal growth time scale so that asymptotic modal growth is not relevant (cf. Fig.~\ref{fig:kcimax}).

In this section the energetics underlying the parametric instability of the Lyapunov vector supporting
S3T turbulence has been studied in detail by analyzing the intricate interplay between the time
dependence of the Lyapunov vector and the time dependence of the mean flow that sustains the
Lyapunov vector by non-normal energetic interaction.

\section{Dynamics of the decaying Lyapunov structures in the presence of  parameterized nonlinear excitation}

Consider S3T dynamics~\eqref{eq:S3T} under stochastic excitation
with covariance $\Qm$ and linear dissipation  at the variable rate ${{\rm Tr}(\Qm)}/{{\rm Tr}(\Cm)}$. This dissipation rate is chosen so that the energy input rate ${{\rm Tr}(\Qm)}$ is equal to the dissipation rate at each time instant so that no net energy is injected into the perturbation field, consistent with the property that the third-order cumulant  being parameterized does not contribute in the  net to the perturbation energy. It is an interesting attribute of even time-independent
non-normal dynamical systems that although this excitation inputs no net energy to the perturbations, still a non-vanishing perturbation field can be sustained by it. This is in contrast to  normal system dynamics,
for in that case the perturbation energy evolution equation,  which is the trace of~\eqref{eq:pS3T}, obeys:
\begin{equation}
\frac{{d} {\rm Tr}( \Cm )}{{d} t}~ = {\rm Tr} \left (  (\Am +\Am^{\dagger} ) \Cm \right )\ , \label{eq:ES3Tf}
\end{equation}
which implies that ${\rm Tr}(\Cm)$ will asymptotically vanish  if $\Am$ is normal and has decaying modes. Such a forcing can sustain a non-vanishing covariance
of substantial rank when $\Am$ is non-normal  even should all modes of the system be damped. That a high-rank  perturbation covariances can be sustained in a turbulent system with a stochastic parameterization of the third cumulant characterized by  zero energy injection has been previously demonstrated  in the context of a discussion of the statistical state dynamics of two-layer baroclinic turbulence~\citep{Farrell-Ioannou-2009-closure}.
A turbulent state with high rank $\Cm$ is also maintained in the S3T turbulence of our Couette flow with an energy neutral parameterization of the third order cumulant, as shown in Fig.~\ref{fig:sigma0f}. In this simulation we have initialized the S3T dynamics with a full rank $\Cm$. The stochastic excitation parameterization of the third cumulant which injects no energy is introduced at $t=200$. In the absence
of excitation the covariance is seen to be in the process of collapsing  to the rank 1
covariance of the first Lyapunov vector with the remaining Lyapunov vectors decaying
at the rate of their respective Lyapunov exponents.
When the excitation is imposed the covariance  rapidly adjusts to maintain a statistically steady state with finite rank
(in this example the rank is approximately 50). The eigenvectors of the finite rank perturbation covariance which are maintained by energy transfer from the fluctuating mean (as shown in  Fig.~\ref{fig:glyap_all}) are called, in analogy to the unforced case, the  stochastically forced Lyapunov vectors (SFLV). These SFLV's inherit the structure of their associated Lyapunov vectors (LV's) as can be seen from Fig.~\ref{fig:lv_sflv} in which the energy norm projections of the LV's
and SFLV's are shown as a contour plot.
Diagonal dominance in this plot indicates that the stochastically maintained Lyapunov vectors are correlated in structure with the underlying Lyapunov vectors which decay in the absence of excitation.

 In summary this section demonstrates, as implied by  the correllation between the LV's and SFLV's shown in  Fig.~\ref{fig:lv_sflv},  that the structure of the perturbation variance in turbulent shear flow is inherited from the Lyapunov vectors, that perturbation variance in shear flow turbulence can be maintained directly by extraction of energy from the mean flow, that the structure of the perturbation field can be predicted to be that of the Lyapunov vectors, and that the contribution of the LV's to the perturbation variance can be ordered in the stability of the LV's.  These implications are corroborated in Fig.~\ref{fig:e_lv_sflv} in which
is shown the perturbation variance fraction accounted for by the SFLV's
(which are identical to the POD modes for the perturbations)  and the LV's ordered in mode number.
The variance is seen to be concentrated in the first few LV's implying that perturbation
structure may be efficiently characterized by making use of these LV structures.

\section{Mechanism regulating the statistical mean state of S3T turbulence}
\label{sec:regulator}

We turn next to study of the  mechanism by which the state of S3T  turbulence is regulated to its observed statistical mean. The observation that the streak is constrained to be marginally unstable (cf. Fig.~\ref{fig:prob_eig} and Fig.~\ref{fig:kcimax}) suggests that the regulation of the turbulent state may be associated with  adjustment to marginal streak stability~\cite{Farrell-Ioannou-2012}. To study the dynamical mechanism regulating the turbulence to a statistical steady state we make use of an analysis of the energetics of the streak (cf. Appendix.~\ref{ap:B}).

\begin{figure*}
\includegraphics[width = .8\textwidth,clip=]{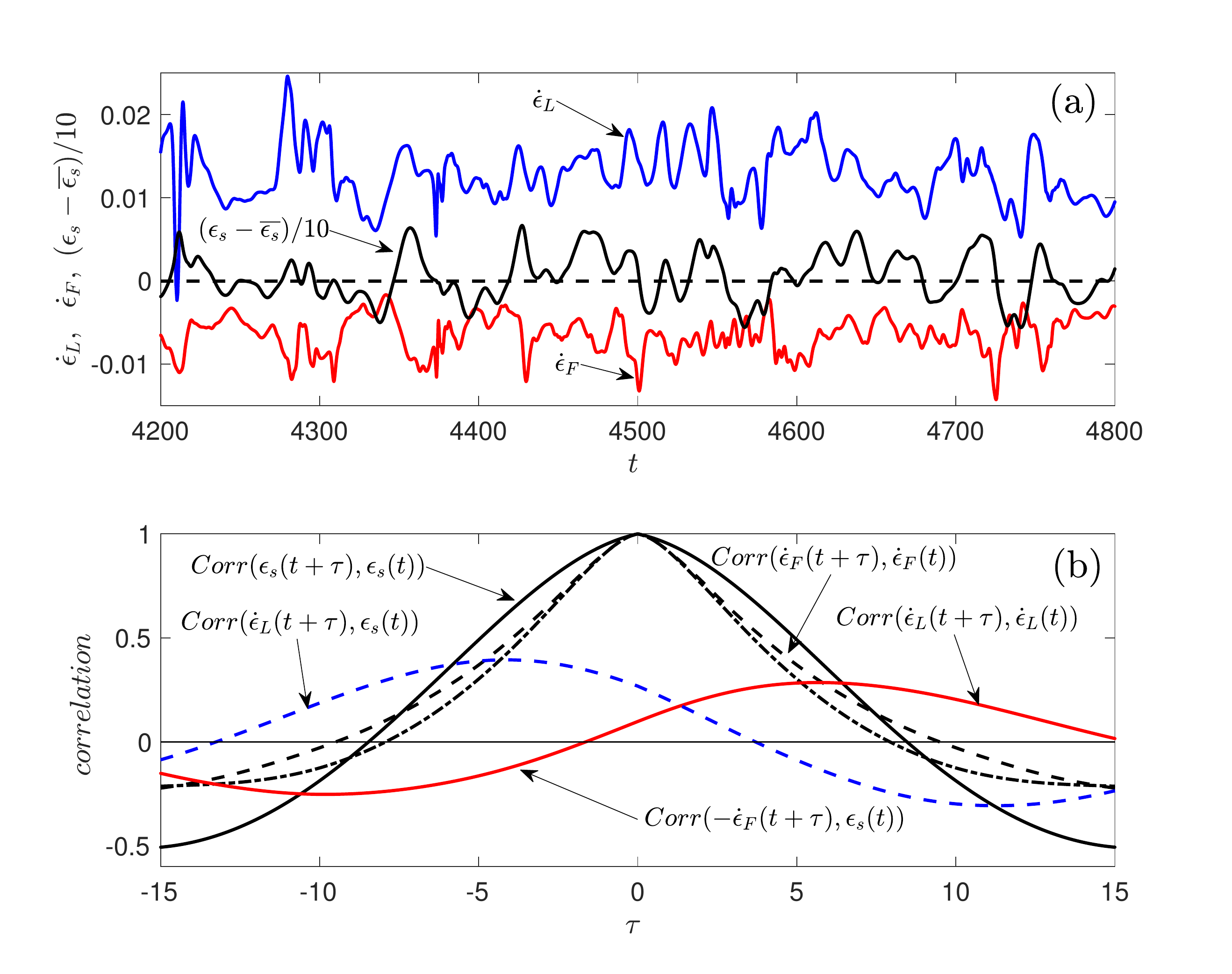}
\caption{\label{fig:regulator}Panel~(a):~Sample time series of the deviations from the time mean of the streak energy, ${\epsilon_s}-\overline {\epsilon_s }$, in an S3T simulation, of the contribution to the time rate of change in streak energy  from mean advection (the lift-up mechanism),  $\dot \epsilon_L$, which is always positive, and of the perturbation Reynolds stress, $\dot \epsilon_F$, which is always negative (cf.~Appendix~\ref{ap:B}). Panel~(b):~Comparison of the autocorrelations of the streak energy $\epsilon_s$, of $\dot \epsilon_F$ and of $\dot \epsilon_L$. Shown also is the cross correlation of these quantities $Corr(-\dot \epsilon_F (t +\tau),\epsilon_s(t))$ and $Corr(\dot \epsilon_L(t +\tau),\epsilon_s(t))$. The cross correlation between $\epsilon_s$ and  $-\dot \epsilon_F$ reveals that these quantities are closely positively correlated with only a $\tau=5$ lead of the streak energy over the Reynolds stress term.}
\end{figure*}

A time series of the Reynolds stress and lift-up  term contributions to the maintenance of the streak energy
are shown in Fig.~\ref{fig:regulator}. Of note is that the Reynolds stress term in the streak energy equation is always negative. In Fig.~\ref{fig:regulator}b is shown the autocorrelation of the streak energy $\epsilon_s$, of the perturbation Reynolds stress
term in the streak energy equation, $\dot \epsilon_F$, and  of the contribution of the lift-up term to maintaining
the streak, $\dot \epsilon_L$, together with
the cross correlation of  $\dot \epsilon_F $ with $\epsilon_s$ and of $\dot \epsilon_L $ with $\epsilon_s$.
The correlation between time series $f(t)$ and $g(t)$ is defined as
\begin{equation}
	Corr(f,g) \equiv  \frac{\overline{ (f-\overline{f}) (g-\overline{g}) }}{\sqrt{\overline{(f-\overline{f})^2}\; \overline{\vphantom{(f-\overline{f})^2}(g-\overline{g})^2}}}~.
	\end{equation}
The cross correlation between $\epsilon_s$ and $-\dot \epsilon_F$ reveals that these quantities are correlated with a $\tau=5$ lead of the streak energy over the Reynolds stress term. This correlation with short lead time in which streak energy maxima are followed by strong Reynolds stress damping indicates a rapidly acting regulation of the streak energy by the Reynolds stress. The small lead time indicates that transient growth on the advective time scale rather than instability growth on the much longer instability time scale $(1/\sigma_{\max}\approx20$) is involved in this regulation of the streak energy (cf.~Fig.~\ref{fig:kcimax}).
Of note is that the lift-up contribution to streak energy leads the streak energy by 5 units of time.

The availability of very rapidly growing  fluxes that damp streak energy and that are
strongly correlated  with streak amplitude explains the robustness of the turbulent state in S3T: the streak grows relentlessly by lift-up due to the roll forcing by the perturbations resulting from the parametric instability of LV$_1$ which would cause the streak amplitude to diverge  were it not for the even stronger transient growth of projections on the adjoint modes associated with incipient streak instability,
which strongly damp the streak energy on the advective time scale, which is short compared to the instability time scale, producing a tightly controlled equilibrium statistical state~\citep{Farrell-Ioannou-2012}.

In summary, this section demonstrates how regulation of S3T turbulence to its statistical steady state is enforced by interaction between the first and second cumulants which completes the analysis of turbulence dynamics in Couette flow at second order in an expansion in cumulants.

\section{Conclusion}\label{sec:conclusion}

The  S3T system is a statistical state dynamics closed at second order that has highly simplified dynamics and naturally self-sustains a turbulent state with restricted support in streamwise wavenumber so that S3T turbulence is dynamically and
computationally an attractive system for studying the mechanism underlying maintenance of wall-turbulence.
S3T system turbulence is in many aspects realistic and in particular it supports a realistic self-sustaining process.
In this work we have exploited the simplicity of the self-sustaining process in S3T turbulence with the stochastic
parameterization of the third cumulant set to zero
to study the mechanisms underlying the maintenance and regulation
of turbulence in this system.
The mechanism maintaining the turbulence is a parametric growth process associated with the time-dependence of the streamwise mean flow streak component and
consistently the  resulting structure of the perturbation state
is that of the first Lyapunov vector supported by the time-dependent streak.
With inclusion of a   stochastic excitation with zero energy injection
parameterizing the perturbation--perturbation nonlinearity
 the perturbation field is supported by the first
Lyapunov vector augmented by the remaining Lyapunov vectors which are induced to extract energy from the mean flow
by the parameterized nonlinearity.  The structure of the incoherent turbulence
perturbations  supported by the  parametric growth process  is shown to remain close to that
of the Lyapunov vectors of the unforced example. Finally, the mechanism by
which the statistical mean state is determined in S3T turbulence is identified to be
a tight balance between robust streak growth by lift-up due to the roll forcing by the
perturbations which in turn results from the parametric instability of the first Lyapunov vector (LV$_1$) and the even
stronger damping resulting from transient growth of the adjoint modes which arise as the streak grows.
These  adjoint modes produce growth that increases rapidly near the stability boundary consistent with
the slight amount of streak instability observed in the simulations~\citep{Farrell-Ioannou-2012}.
These competing processes of robust streak
growth opposed by strong damping produce a tightly controlled equilibrium statistical state.


\appendix 
\section{Lyapunov exponents and vectors}
\label{ap:A}

Consider the time dependent linear dynamical system:
\begin{equation}
	\dot x = \Am(t) x\ ,
\end{equation}
with $x$ an $n$ dimensional state vector and $\Am$ a bounded  $n \times n$ time dependent matrix. If the state of the system  at time $t_0$ is $x(t_0)$,  the state of the system at time $t$ is given by
\begin{equation}
	x(t) = \Phim (t,t_0) \, x(t_0)~,
\end{equation}
where the propagator, $\Phim(t,t_0)$, is the $n \times n$ matrix that maps the state vector at time $t_0$ to the state vector at time $t$.

 The Lyapunov exponents are
defined to be the various limits
\begin{equation}
\lambda = \lim_{t\rightarrow \infty} \frac{\log \| \Phim (t,t_0)\, x(t_0)\|}{(t-t_0)}~,
\end{equation}
 that can occur as $x_0$ spans the space of all possible initial conditions. We denote with $\| \bcdot \|$ the norm chosen to measure the vector magnitude. The Lyapunov exponents are  norm independent
 and also independent
 of the initial time $t_0$.
 Oseledets's theorem~\cite{Oseledets-1968} guarantees that there are $n$ such Lyapunov exponents $\lambda_1 > \lambda_2 > \dotsb  > \lambda_n$
 (under the assumption that there is no degeneracy in the  values of the Lyapunov exponents)  that can be  obtained as eigenvalues of the Hermitian positive matrix:
 \begin{equation}
\Lm_{\infty}(t)=\lim_{t_0 \to -\infty} \frac{\log \left ( \Phim (t, t_0) \Phim^\dagger(t,t_0) \right )}{2(t-t_0)}~.
\end{equation}
In the above definition the inner product is taken to be the dot product  which is natural in our examples as our variables are velocities so the dot product results in a norm proportional to energy.
We refer to the orthogonal time dependent eigenvectors $u_1(t),u_2(t),\dots,u_n(t)$ of $\Lm_{\infty}(t)$ as the Lyapunov vectors (LVs) of the system. With the exception of the first, these vectors  depend on the chosen inner product consistent with their being orthogonal in that inner product (energy in our case).
The time dependent eigenvector, $u_1(t)$,  corresponding to the  maximal Lyapunov
exponent, $\lambda_1$, is called the first Lyapunov vector, LV$_1$.

Vectors proportional to  LV$_1$, form subspace $E_1(t)$,  and their magnitude changes with time as $t \to -\infty$ as $\exp(\lambda_1 t)$. LV$_1$ therefore becomes the dominant  structure  after a sufficiently long integration of the system (assuming no degeneracy of the first Lyapunov exponent).   Vectors in the subspace $E_2(t)$
spanned by $u_1(t)$ and $u_2(t)$, except those that are proportional to $u_1(t)$,  decay as $t \to -\infty$ as $\exp(\lambda_2 t)$ and $u_2(t)$ is referred to as the second Lyapunov vector, LV$_2$. In this way the state space is split
into a set of  nested subspaces $E_1(t) \subset E_2(t) \subset \dotsb \subset E_n(t)$ such that
the vectors that are in $E_i(t)$ and are not in subspace $E_{i-1}(t)$ decay  as $t \to -\infty$ as $\exp(\lambda_i t)$.
 This definition of the Lyapunov vectors  was introduced by
Lorenz~\cite{Lorenz-1984} in his studies of error growth in atmospheric dynamics; see also Farrell \& Ioannou~\cite{Farrell-Ioannou-1996b} and Wolfe \& Samelson~\cite{Wolfe-Samelson-2007}.
Assuming a physically based inner product, e.g.  perturbation energy in our case, the orthogonal basis defined by these Lyapunov vectors provides
a physically meaningful orthogonal basis
for partitioning  the state space of the evolving perturbations in the sense that perturbation states  that were in the far past in a  sphere of unit energy will evolve at time $t$ into
an ellipsoid  the principle axes of which lie in the direction of the  LV's and  partition the state space into subspaces spanned by these vectors which are ranked in magnitude in the order of their Lyapunov exponents as $\exp(\lambda_i t)$.

It should be noted that the Lyapunov vector $u_i(t)$, with $i>1$, when integrated forward will not in general grow asymptotically at rate
$\lambda_i$ (but almost surely at rate
$\lambda_1$).  This fact has two equally important roots.  The first is mathematical: because of the orthogonalization procedure imposed on the Lyapunov vectors at each time step the components of the temporally evolving state vector growing at the rate of $\lambda_i$ that lie in directions spanned by  previous Lyapunov vectors LV$_1$ through $LV_{i-1}$ is being projected out.  While the orthogonal LV decomposition retains information on the subspace spanned by the Lyapunov vectors, it results in loss of the information on which structures are growing at the rate of each Lyapunov exponent at each time, with the exception of the first~\cite{Yang-Radons-2010}.   The second root is more physically relevant: from a physical perspective this follows from the fact that a random vector perturbation has measure zero probability of having zero projection on LV$_1$ and so any random perturbation results in growth that is   asymptotically at rate $\lambda_1$.  Besides being of profound physical significance, this universal property of all physical vectors asymptotically converging to LV$_1$ poses a problem for calculation of the Lyapunov vectors.  In order to obtain the Lyapunov vectors operationally at all times we integrate forward the time dependent Lyapunov equation for the covariance~\eqref{eqn:p2} and after a sufficiently  long integration the Lyapunov vectors at time $t$ emerge as the eigenvectors of the covariance matrix
$\Cm(t)$. The eigenvectors of $\Cm(t)$ define the Lyapunov vectors that are orthogonal in the energy inner product.

In some recent studies calculations were performed to determine at each time $t$ the vectors that grow when integrated forward and decay when integrated backwards at the rate of the  corresponding
Lyapunov exponent~\cite{Cvitanovic-etal-2016}. These vectors, called confluent Lyapunov vectors (CLV),
generalize to time dependent linear systems the eigenvector analysis of time independent linear systems~\cite{Ginelli-etal-2007,Wolfe-Samelson-2007,Yang-Radons-2012}.
However, the confluent Lyapunov vectors are not orthogonal in any physical norm.   In order to use them to partition energy growth as we do in our analysis  the additional step of orthogonalizing the CLV's in energy would have to be performed, which would serve to recover the  LV's that  we have discussed.

\section{Streak Energetics} \label{ap:B}

%
%
%
The streak component of the mean streamwise velocity, $U$,
is defined as $U_{\rm s} = U- [ U ]_z$, where $[\,\bcdot\,]_z$ denotes the spanwise average.
The streak is the part of the streamwise  velocity  with  zero $x$ wavenumber but nonzero  $z$ wavenumber Fourier components.
By subtracting~\eqref{eq:MU} from  its spanwise average we obtain an equation for the evolution of the streak velocity:
\begin{align}
\partial_t U_{\rm s} & =- \partial_y\( U V - [UV ]_z \)- \partial_z\( U W \) \nonumber\\
&   -\partial_y \(
[uv]_x-[uv]_{x,z} \) - \partial_z \(
[uw]_x\) + \Delta U_{\rm s}/R~,\label{eq:usbar}
\end{align}
using notation~\eqref{eq:x}, and from~\eqref{eq:usbar} we obtain  the following evolution equation for the streak energy, $\epsilon_{s} \equiv \int_{-1}^1  \left[ U_s^2/2 \right]_z \, \df y$:
\begin{equation}
	\dot \epsilon_s = \dot \epsilon_L + \dot \epsilon_F + \dot \epsilon_D\ ,
\end{equation}
where
\begin{equation}
 \dot \epsilon_L =   -\int_{-1}^1 \left [ \bit U_{\rm s} \, (  V\partial_y U+W\partial_z U ) \right ]_z  \df y \, ,\end{equation}
is the contribution to the streak energy rate of growth from  advection of mean $U$ momentum by
the $V$ and $W$ velocities.
The term
  \begin{equation}
    \dot \epsilon_F =  -\int_{-1}^{1} \,\[  U_{\rm s} \partial_y \(  [u v]_x \) + U_{\rm s} \partial_z \( [u w]_x \)\bit \]_z  \df y \,  ,
  \end{equation}
  is the contribution to the streak energy rate of growth from the perturbation  Reynolds-stress divergence,  and
  \begin{equation}
  \dot \epsilon_D = \frac1{R}\int _{-1}^1 \, \left [  U_{\rm s}\,\Delta U_{\rm s} \right ]_z\, \df y\ ,
  \label{eq:energUd}
\end{equation}
is the rate of dissipation of streak energy.

\begin{acknowledgments}
This work was initiated during the 2013 First Multiflow Summer Workshop at  the Universidad Polit\'{e}cnica de Madrid  with financial support from the Multiflow Program of the European Research Council. We would like to thank Prof.~Javier Jim\'{e}nez, Prof.~Dennice Gayme,
Dr.~Vaughan Thomas, Dr.~Adri\'an Lozano-Dur\'an and Dr.~Navid  Constantinou
for useful comments and fruitful discussions. Brian Farrell was
 partially supported by  NSF AGS-1246929.
\end{acknowledgments}


%

\end{document}